\documentclass[prl,twocolumn,amsmath,amssymb]{revtex4}
\begin{document}
\author{Kevin Leung,$^{1*}$ Ida M.B.~Nielsen,$^2$ and Louise J.~Criscenti$^1$}
\affiliation{$^1$Sandia National Laboratories, MS 1415, \& 1322,
Albuquerque, NM 87185\\
$^2$Sandia National Laboratories, MS 9158, Livermore, CA 94551\\
$^*${\tt kleung@sandia.gov}}
\date{\today}
\title{Elucidating the bimodal acid-base behavior of the water-silica
interface from first principles}

\input epsf
%\ssp
                                                                                
\begin{abstract}
 
Understanding the acid-base behavior of silica surfaces is critical for
many nanoscience and bio-nano interface applications.  Silanol groups
(SiOH) on silica surfaces exhibit two acidity constants---one as acidic
as vinegar---but their structural basis remains controversial.  The
atomic details of the more acidic silanol site govern not just the
overall surface charge density at near neutral solution pH,
but also how ions and bio-molecules interacts with and bind to silica
immersed in water.  Using {\it ab initio} molecular dynamics
simulations and multiple representative crystalline silica surfaces,
we determine the deprotonation free energies of silanol groups
with different structural motifs.  We show that previously
proposed motifs related to chemical connectivity or inter-silanol
hydrogen bonds do not yield high acidity.  Instead, a plausible candiate
for pK$_{\rm a}$=4.5 silanol groups may be found in locally strained
or defected regions with sparse silanol coverage.  In the process, 
irreversible ring-opening reactions of strained silica
trimer rings in contact with liquid water are observed.  
 
\end{abstract}
                                                                                
\maketitle

\section{Introduction}
                                                                                
Deprotonation of silanol (SiOH) groups\cite{book,iler} at water-silica
interfaces is one of the most common and important, yet intriguing,
interfacial chemical reactions.  Silica (SiO$_2$) is a major component of rocks
and lines the channels of many nanofluidic devices.\cite{dekker1,baca,yang}
Deprotonation governs dissolution rates,\cite{book} affects lipid binding
to silica nanostructures,\cite{baca} creates negative surface charges that
can be tuned with moderate changes in solution pH to perform desalination
and ion
gating,\cite{yang} and may even hinder extraction of positively charged
crude oil components\cite{oil} from underground deposits.  In particular,
the atomic level structural details of deprotonated SiOH groups govern both
the overall surface charge density and the binding of ions and molecules
to immersed silica surfaces.  In this work, we apply {\it ab initio}
molecular dynamics\cite{cpmd} (AIMD), which takes into account proton
dynamics and hydrogen bond network fluctuations in liquid water essential
to acid-base reactions in small molecules\cite{sprik,klein,parrin1,chandler}
as well as cooperative hydroxyl hydrogen bonding behavior specific to oxide
surfaces,\cite{hass,pore} to investigate the enigmatic SiOH deprotonation
equilibrium constant as a function of structural motifs.
 
Measurements of interfacial pK$_{\rm a}$ (defined as $-\log_{10}$ K$_{\rm a}$,
where K$_{\rm a}$ is the acid dissociation constant) have been revolutionized
by surface-sensitive second harmonic generation (SHG) and sum frequency
vibrational spectroscopy (SFVS) techniques.\cite{eisenthal1,shen1}
In 1992,  Ong {\it et al}.\cite{ong} demonstrated that 19\% of silanol
groups on fused silica surfaces exhibit a pK$_{\rm a}$ of 4.5,
about the same as vinegar (acetic acid), while 81\% exhibit pK$_{\rm a}$=8.5.
SFVS experiments on $\alpha$-quartz reached similar conclusions and
further suggested that the low-acidity silanol groups reside in regions
with strong water-water hydrogen bonds.\cite{shen05} A titration study on
silica gel (amorphous silica)\cite{allen} and X-ray photoelectron
spectroscopy measurements on quartz\cite{duval} also independently
demonstrated the existence of SiOH groups with pK$_{\rm a}$ between 4 and
5.5.  Such qualitative agreement on different forms of silica
is expected because liquid water is known to react with even crystalline
silica to form an amorphous layer.\cite{iler,shultz} These measurements
suggest that the earlier, single pK$_{\rm a} \sim $6.8 reported in
amorphous silica titration experiments\cite{schindler} may reflect
a composite of two types of SiOH.
 
\begin{figure}
\centerline{\hbox{(a)\epsfxsize=1.50in \epsfbox{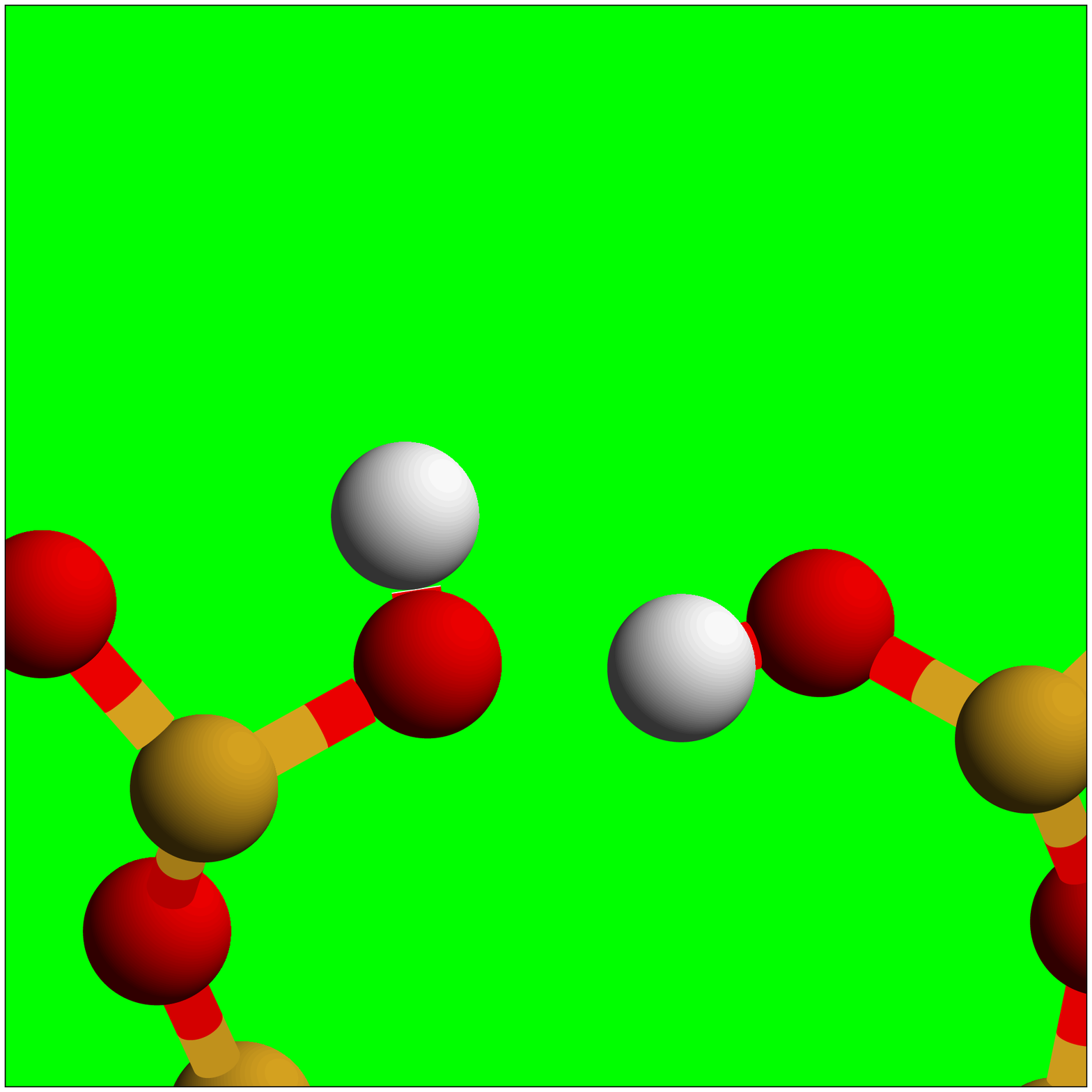} }
                    {\epsfxsize=1.50in \epsfbox{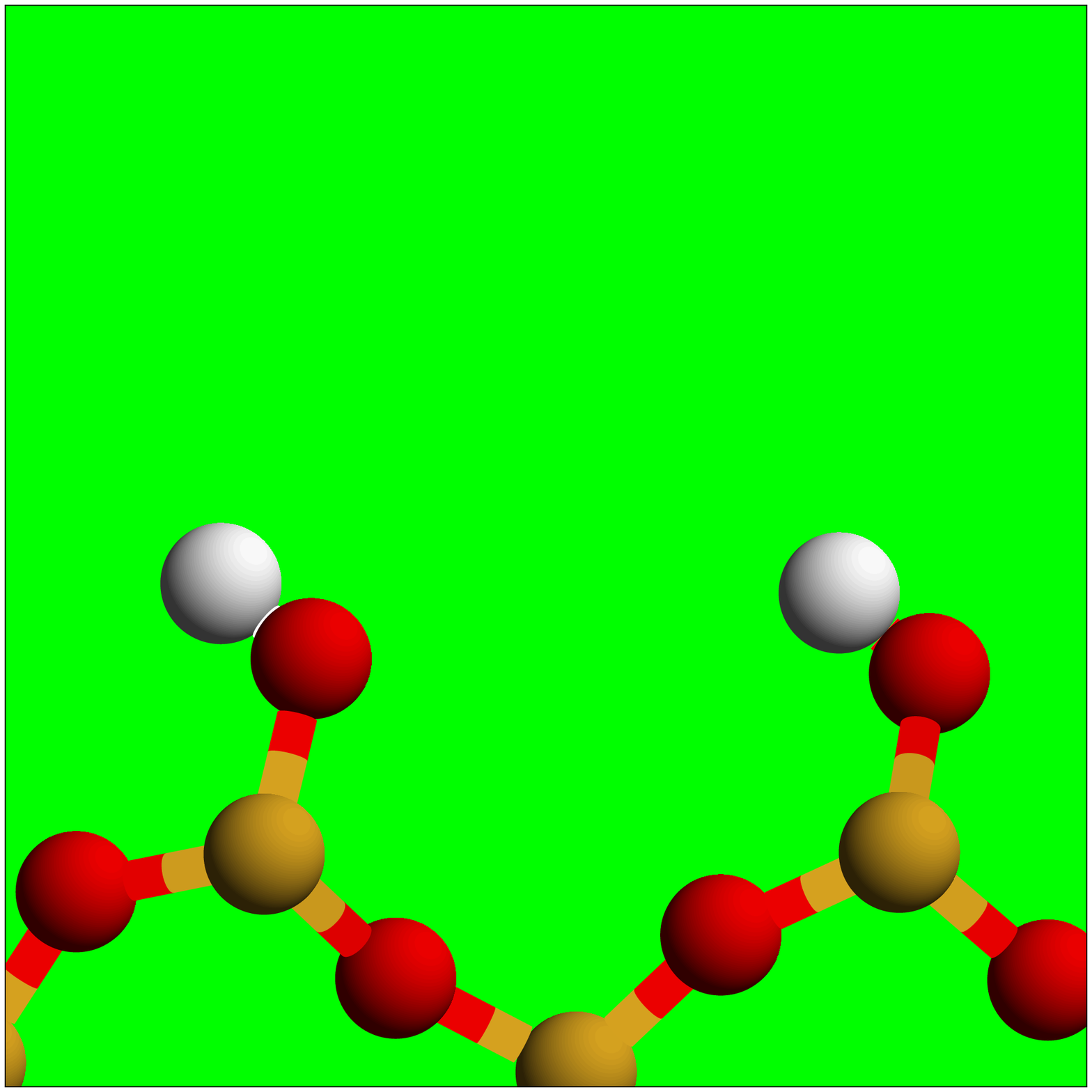} (b)}}
\centerline{\hbox{(c)\epsfxsize=1.50in \epsfbox{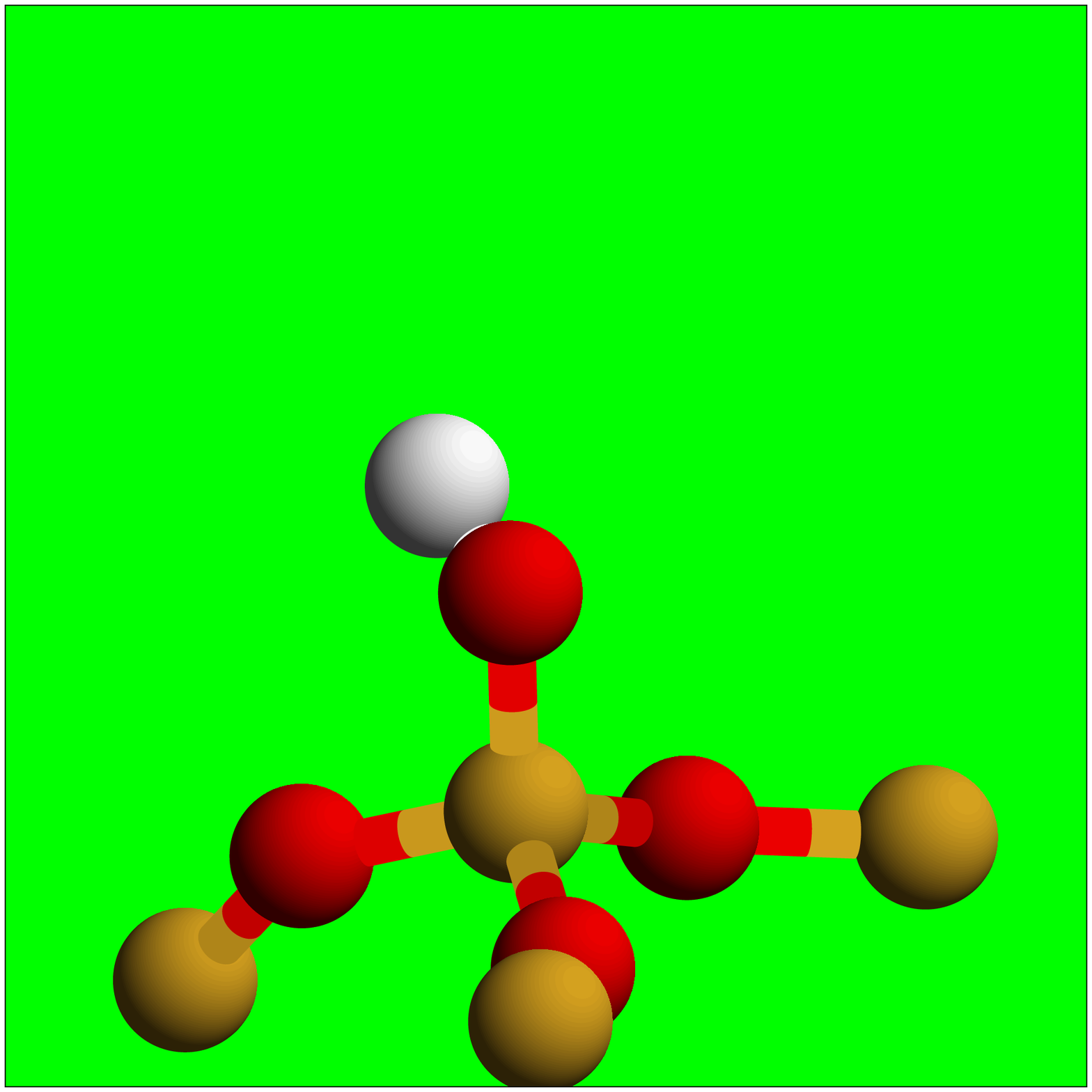} }
                    {\epsfxsize=1.50in \epsfbox{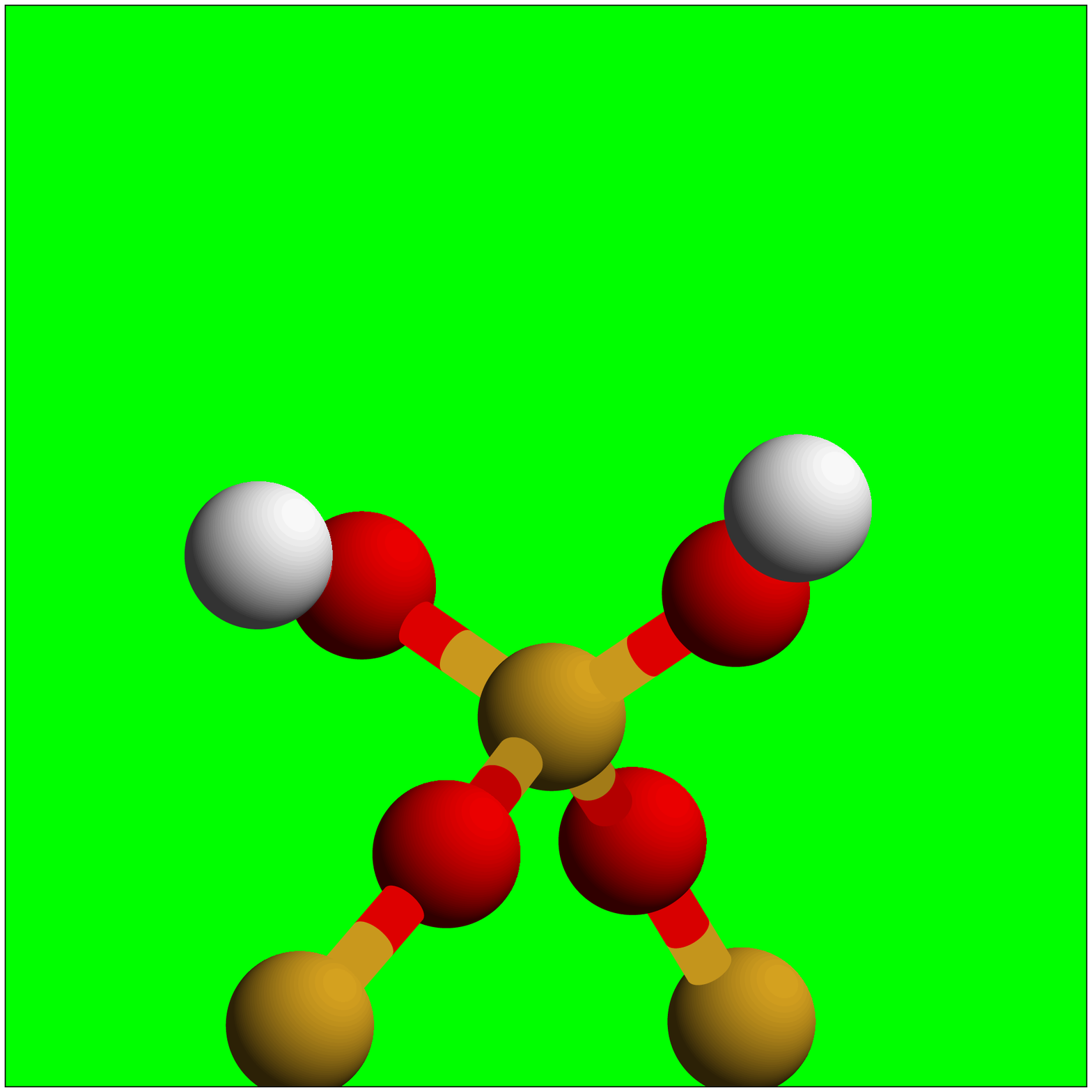} (d)}}
\caption[]
{\label{fig1} \noindent
(a)-(d) Four types of SiOH groups discussed in this work.
(a) Hydrogen-bonded; (b) isolated; (c) Q$^3$; (d) Q$^2$.
Oxygen, silicon, and hydrogen atoms are colored in yellow, red,
and white, respectively.
}
\end{figure}

The acidities of surface silanol groups have been assigned to 
different chemical connectivities or inter-silanol hydrogen bonding.
In accordance with the literature, we differentiate SiOH groups
according to whether they are directly hydrogen bond to other
SiOH (``H-bonded,'' Fig.~\ref{fig1}a), or are not so hydrogen-bonded
(``isolated,'' Fig.~\ref{fig1}b); whether the 4-coordinated Si atom of
the SiOH is part of 3 covalent Si-O-Si- linkages (``Q$^3$,'' Fig.~\ref{fig1}c),
or only part of 2 Si-O-Si (``Q$^2$,'' Fig.~\ref{fig1}d).  It has
been suggested that the ratio of H-bonded to isolated SiOH is
about 1 to 4, similar to the relative occurrence of pK$_{\rm a}$=4.5 and
8.5;\cite{ong} thus pK$_{\rm a}$=4.5 has been
ascribed to isolated silanol groups.\cite{ong,dong,lorenz,fan1}
On the other hand, the Q$^2$:Q$^3$ ratio has also been described
as either approximately 1:4 or 4:1, which has prompted assignment
of the pK$_{\rm a}$=4.5 SiOH group to either Q$^2$
(Refs.~\onlinecite{mori1,nanosilica})
or Q$^3$ (Refs.~\onlinecite{others,shaw2,rosenholm}).  
The conflicting estimates of the ratios of silanol groups with
different structural motifs\cite{mori1,nanosilica,others,shaw2,rosenholm}
likely reflect the difficulty brought about by the tendency of
liquid water to react with crystalline silica.\cite{shultz}
One indisputable experimental finding is that the silanol surface density is
$\sigma_{\rm SiOH}\approx$ 4.6~nm$^{-2}$ on well-soaked amorphous
samples.\cite{iler,zhuravlev} On the
theoretical side, static geochemical models,\cite{hiemstra,bickmore} which
do not account for aqueous phase hydrogen bonding and dynamical proton
motion, have also been applied, but they have not yet explained the two
observed acidity constants,\cite{ong,shen05,allen} while quantum chemistry or
DFT methods with a dielectric continuum treatment of the bulk water
environment have been limited to calculating the pK$_{\rm a}$ of small
silica fragments.\cite{sahai2,rustad1,sefcik}  DFT modeling of
amorphous silica slabs have also been considered,\cite{mauri} but the
pK$_{\rm a}$ estimates therein often do not treat water explicitly or
dynamically.  (See Supporting Information (SI) for discussions
of the significance of hydrogen-bond network fluctuations and excess proton
hopping.)
 
\begin{figure}
\centerline{\hbox{ (a) \hspace*{1.35in} (b) \hspace*{1.35in} (c)}}
\centerline{\hbox{\epsfxsize=1.00in \epsfbox{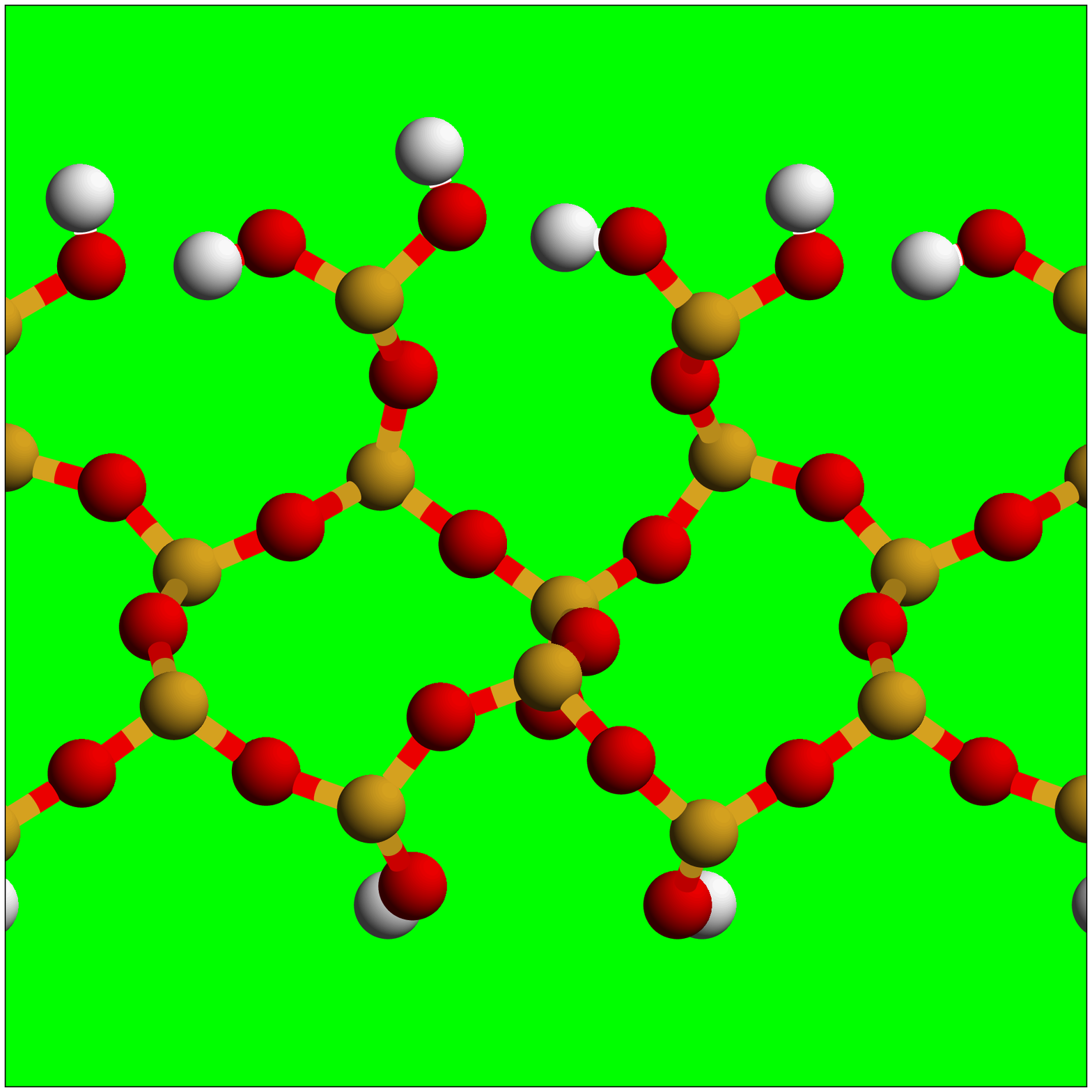} }
                 {\epsfxsize=1.00in \epsfbox{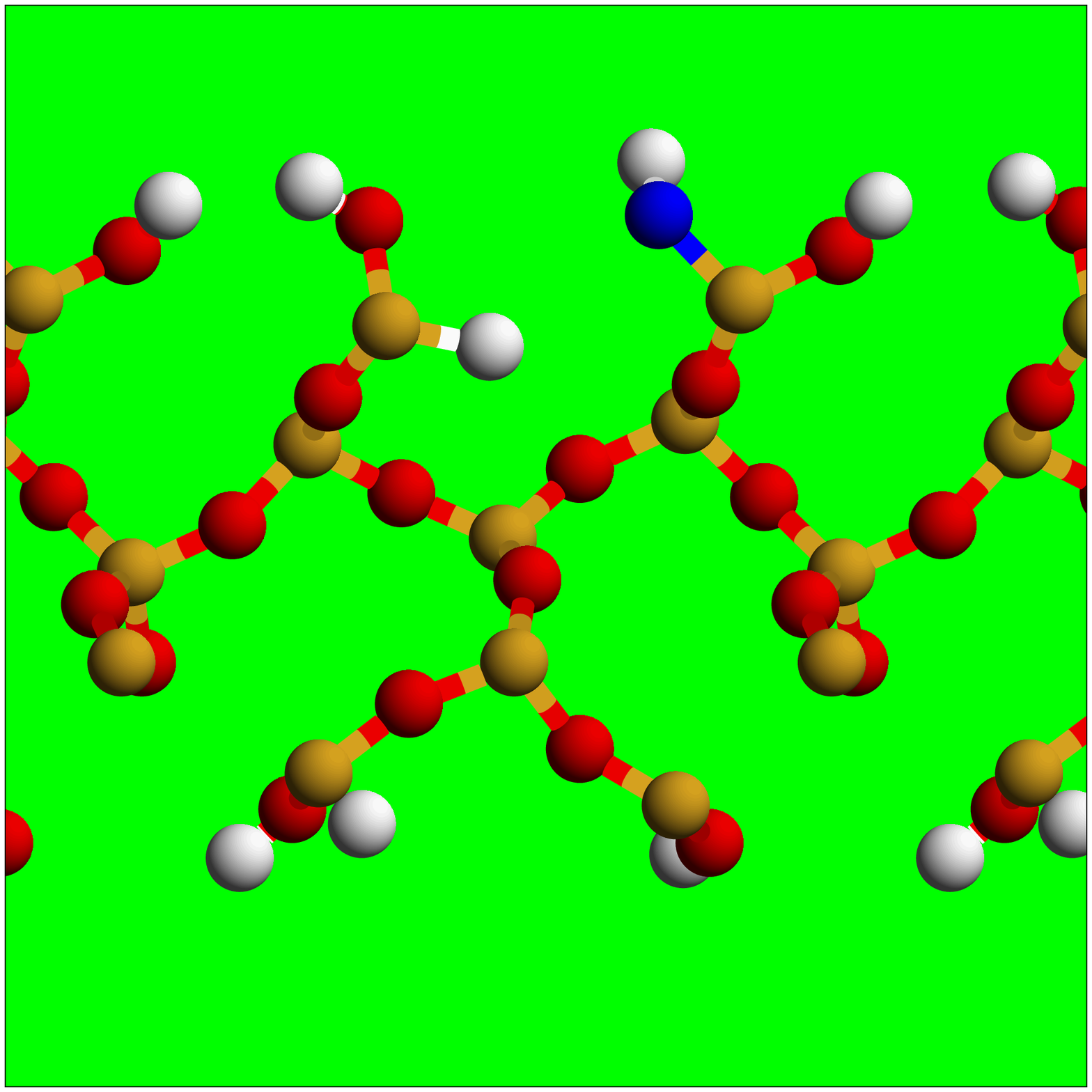} }
                 {\epsfxsize=1.00in \epsfbox{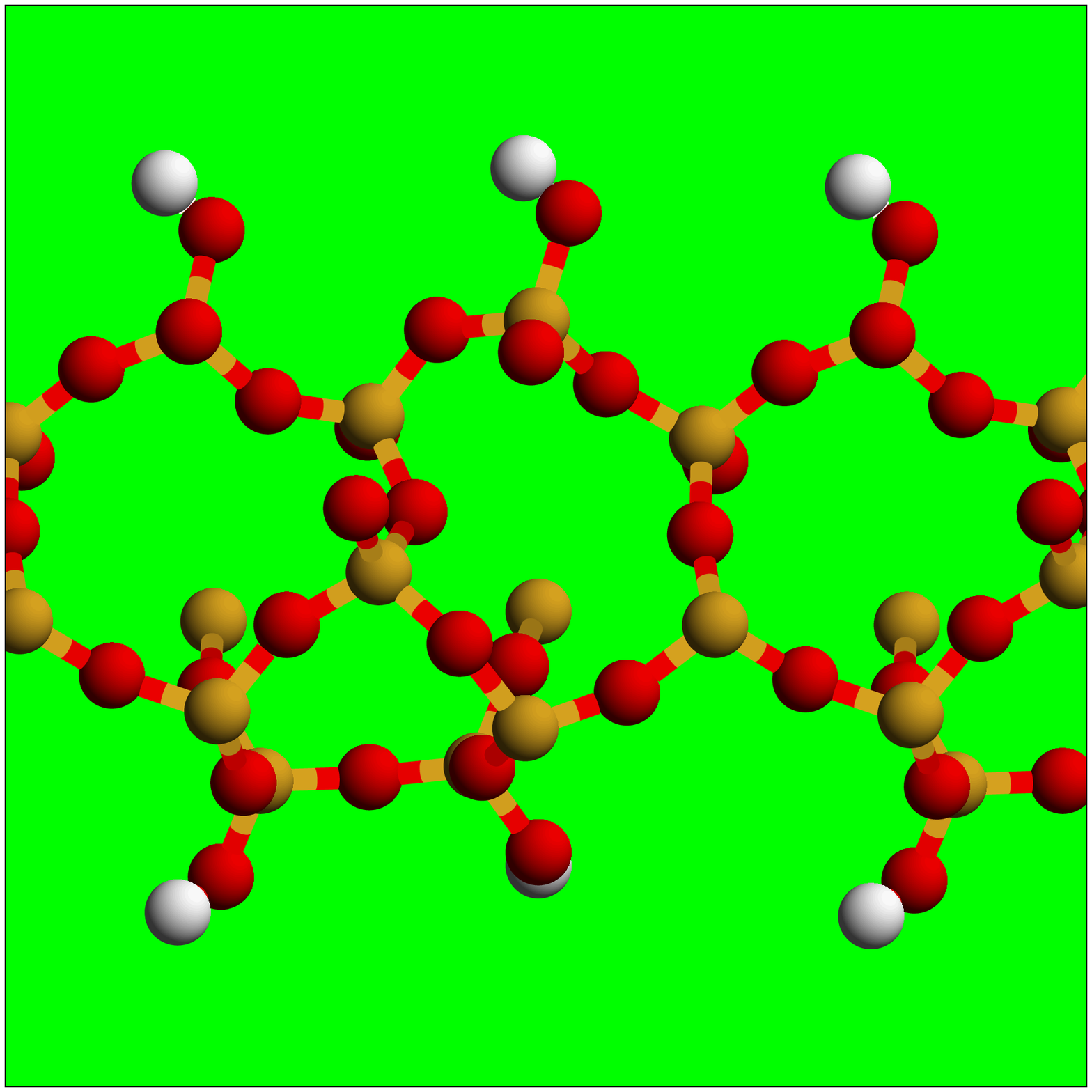} }}
\centerline{\hbox{ (d) \hspace*{1.35in} (e) \hspace*{1.35in} (f)}}
\centerline{\hbox{\epsfxsize=1.00in \epsfbox{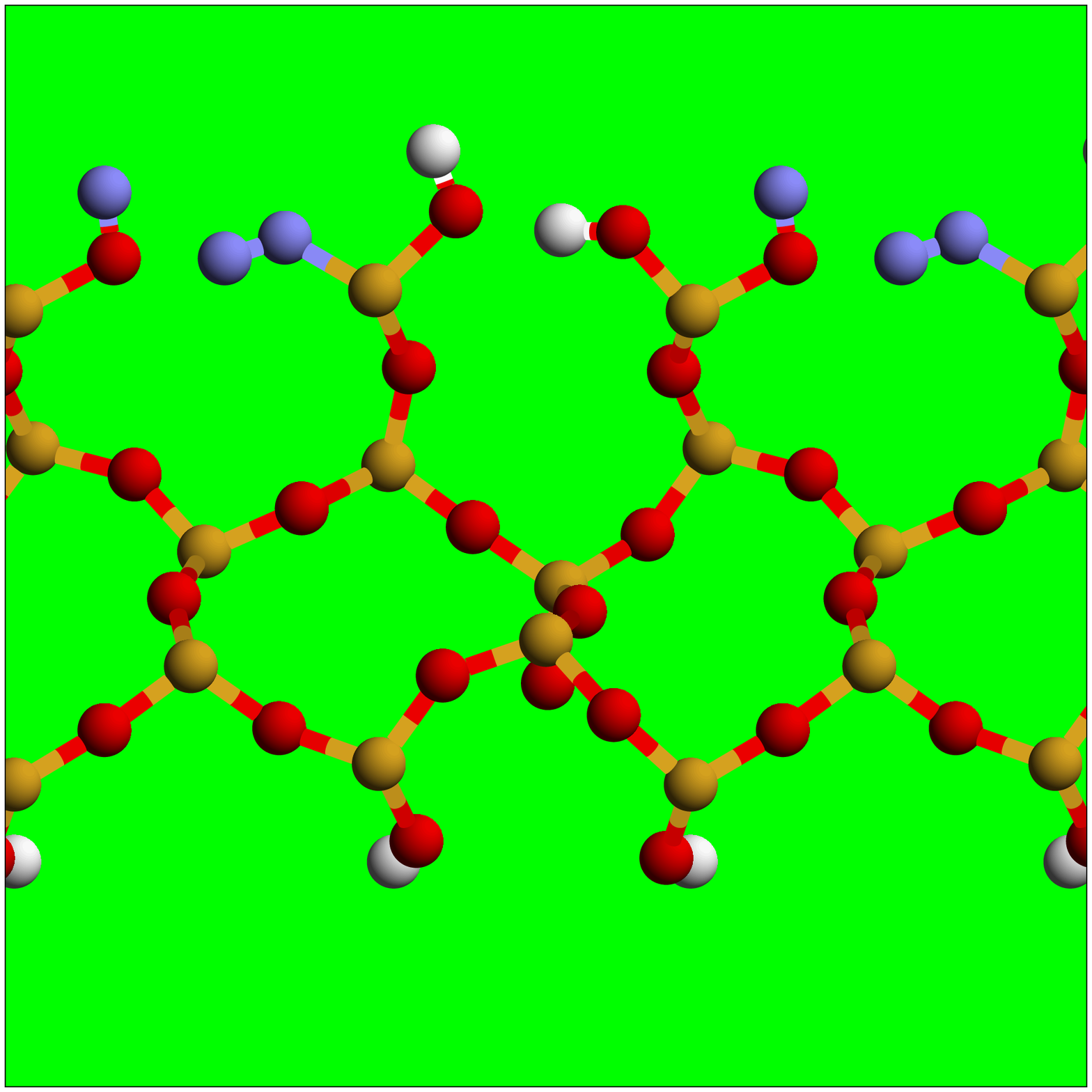} }
                 {\epsfxsize=1.00in \epsfbox{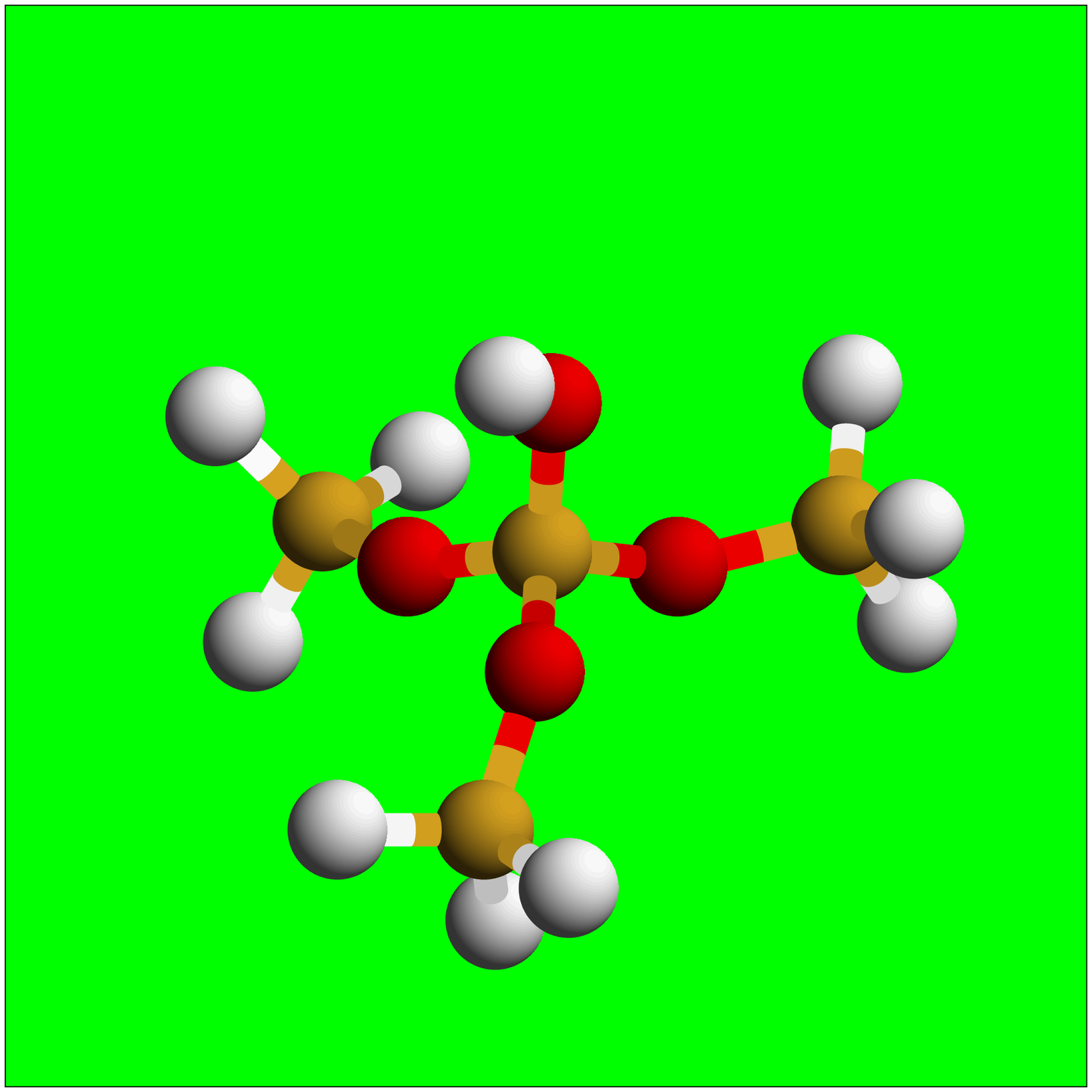} }
                 {\epsfxsize=1.00in \epsfbox{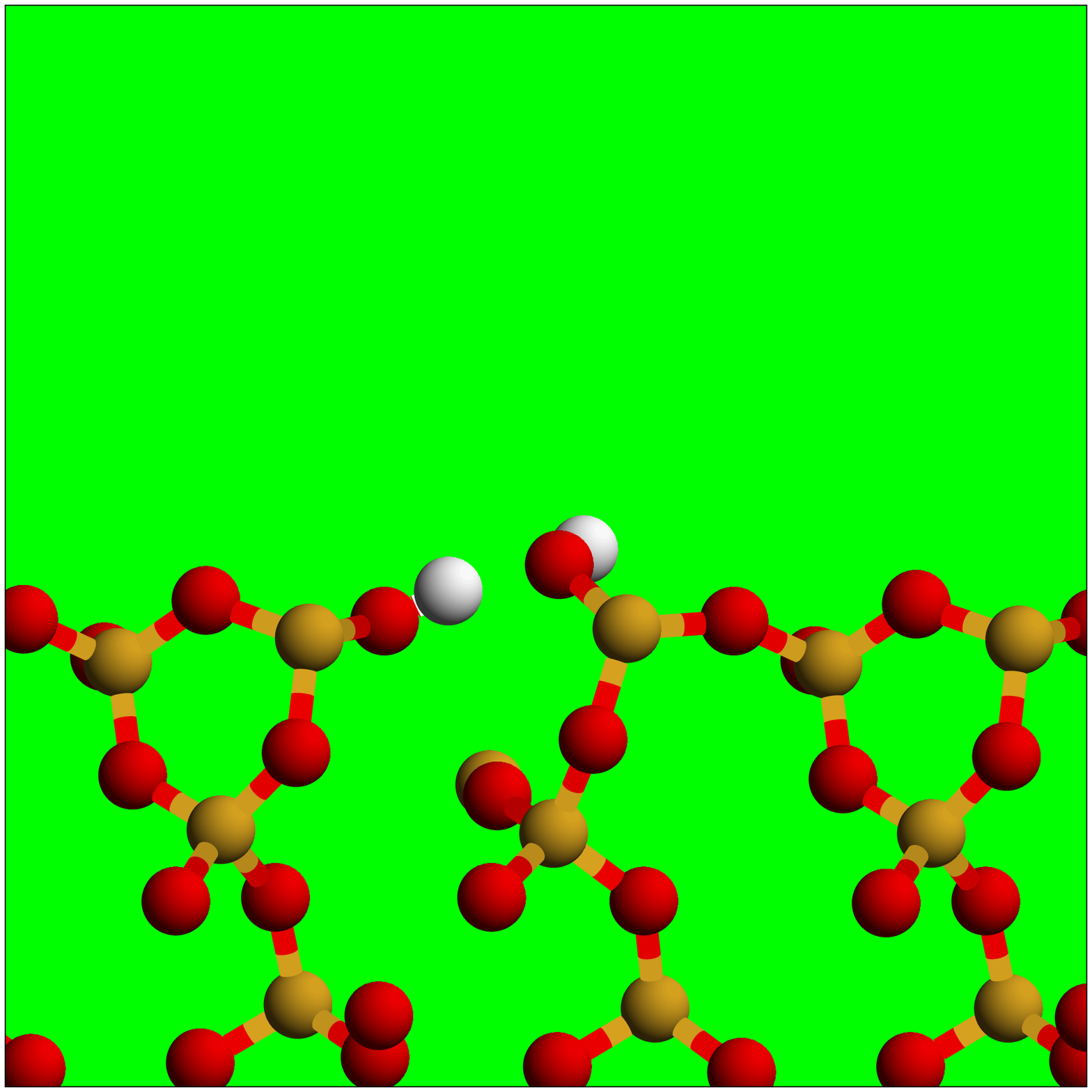} }}
\caption[]
{\label{fig2} \noindent
The heterogeneous SiOH environments examined in this work.
(a) Hydroxylated $\beta$-cristobalite (100) surface.  The SiOH groups have
$\sigma_{\rm SiOH} \sim 8$ nm$^{-2}$, are Q$^2$ and H-bonded.
pK$_{\rm a}$=$7.6~\pm 0.3$.
(b) Hydroxylated $\beta$-cristobalite (100) surface with one SiOH group replaced
with a SiH to break the chain of hydrogen bonds.  The tagged (deep blue)
SiOH (Q$^2$ and isolated) exhibits pK$_{\rm a}$=8.9$\pm 0.3$.
(c) Reconstructed $\beta$-cristobalite (100) surface, 
$\sigma_{\rm SiOH} \sim 4$ nm$^{-2}$, Q$^3$ and isolated;
pK$_{\rm a}$=8.1$\pm 0.5$ (6 layers of water) and~7.0$\pm 0.4$ (4 layers).
(d) The structure in panel (c) comes from removing atoms shown here in 
blue, and attaching the resulting undercoordinated Si and O atoms.
Panels (a)-(d) represent $\sim $1.5 simulation cells in the lateral direction.
(e) (H$_3$SiO)$_3$SiOH, which is Q$^3$ and isolated, exhibits
pK$_{\rm a}$=7.9$\pm 0.5$.
(f) Top half of a reconstructed quartz (0001)
surface model containing cyclic silica trimers (Si-O)$_3$,
$\sigma_{\rm SiOH} \sim 2.3$ nm$^{-2}$, Q$^3$ and H-bonded; one SiOH
resides on a trimer ring (pK$_{\rm a}$=5.1$\pm 0.3$), the other does not
(pK$_{\rm a}$=3.8$\pm 0.4$ and pK$_{\rm a}$=4.8$\pm 0.4$ depending
on whether a nearby trimer ring breaks; see text).  Si, O, and H atoms
are in yellow, red, and white, respectively. (b) and (e) are
finite-temperature AIMD snapshots, with water molecules
omitted for clarity, while (a), (c), and (f) are shown at T=0~K.
}
\end{figure}

AIMD simulations have successfully reproduced the pK$_{\rm a}$ of molecules
in aqueous solution\cite{sprik,klein,parrin1} and should be particularly
well-suited for distinguishing {\it relative} SiOH pK$_{\rm a}$ that are
4 pH units apart in different environments, provided we can demonstrate
that reproducible pK$_{\rm a}$ for chemically equivalent SiOH's can be
predicted.  Coupled with static high-level quantum chemistry corrections,
they provide the most rigorous predictions for liquid state reactions.  As
computing power has increased, AIMD modeling of liquid water-material
interfaces has become viable,\cite{hass,pore,marx,galli2,mundy,car,angelo}
although it remains costly because water dynamics is slower at
interfaces.\cite{funel,rossky}  Furthermore, investigation of
multiple reaction sites and/or crystalline facets is often necessary when
dealing with material surfaces.  In this work, we study the
bimodal acid-base behavior of silanol groups\cite{ong,shen05,allen}
by performing AIMD simulations to
directly calculate the pK$_{\rm a}$ value.  Given the absence of
well-defined water-crystalline silica interfaces,\cite{shultz} and
the fact that the precise atomic structure of amorphous silica surfaces
is unknown, we examine six distinct, representative silanol
environments.  These include hydroxylated $\beta$-crystobalite (100)
(Fig.~\ref{fig2}a), hydroxylated $\beta$-cristobalite (100) with one SiOH
removed (Fig.~\ref{fig2}b), reconstructed $\beta$-cristobalite (100)
(Fig.~\ref{fig2}c), a molecular system (Fig.~\ref{fig2}e), and
two distinct SiOH on reconstructed quartz (0001) (Fig.~\ref{fig2}f).
They represent the SiOH motifs proposed to be responsible for
pK$_{\rm a}$=4.5 or 8.5 in the literature (Figs.~\ref{fig1}a-d).
 
\section{Computational Methods}

AIMD simulations apply the Perdew-Burke-Ernezhof (PBE)
functional,\cite{pbe} the Vienna Atomic Simulation Package
(VASP),\cite{vasp,vasp1} a 400~eV energy cutoff, $\Gamma$-point
sampling of the Brillouin zone, deuterium mass for all protons,
and a 0.375~fs time step at each Born-Oppenheimer dynamics time step.
The trajectories are thermostat at T=425~K; elevated temperature is
needed to represent liquid water properties when the PBE functional
is applied and quantum nuclear effects are neglected, which is the 
case herein (SI, Sec.~S1).  Four to six umbrella sampling windows
of 20~ps production trajectory length each are used per pK$_{\rm a}$
calculation on hydroxylated $\beta$-cristobalite (100)
(Figs.~\ref{fig2}a,b) and reconstructed $\beta$-cristobalite
(Fig.~\ref{fig2}c) surfaces.  These periodically replicated simulation
cells measure 10.17$\times$10.17$\times$26 \AA$^3$.  Their lateral
dimensions are commensurate with simulation cells often used for AIMD
studies of pure water structure or ion hydration.  This is one of the
reasons we have focused on the (100) rather than the (111) surface of
$\beta$-cristobalite, which actually has a $\sigma_{\rm SiOH}$ similar
to that of amorphous silica.\cite{iler}  The more
computationally costly reconstructed
quartz (0001) system has a cell size of 10.0$\times$17.32$\times$24 \AA$^3$,
and 12 to 18~ps trajectories are used per window.  
For each crystalline silica simulation cell, we always start from crystal
slab structures optimized at zero temperature using Density Functional
Theory (DFT) and the PBE functional.\cite{pbe}  Next we switch to the
CHARMM SiO$_2$ force field\cite{charmm} and the SPC/E model for
water.\cite{spce}  The number of water molecules occupying the simulation
cell is determined using these force fields, the Grand Canonical Monte
Carlo (GCMC) technique, and the Towhee code.\cite{gcmc}  With this approach,
the spaces between hydroxylated and reconstructed $\beta$-cristobalite
surfaces are filled with 58 and 63 water molecules, respectively, which
amount to roughly six layers of water, while the reconstructed
quartz simulation cell contains about 63 (about 4 layers of) water molecules.
The (H$_3$SiO)$_3$SiOH pK$_{\rm a}$ calculation utilizes a ($12.42$ \AA)$^3$
cell with 57~H$_2$O and 20~ps sampling trajectories.  Finally, to
check system size dependences, umbrella sampling simulations for a
smaller reconstructed $\beta$-cristobalite (100) simulation cell, 20~\AA\,
in the $z$-direction and containing 44 (4 layers of) water molecules, are
also conducted for at least 10~ps per window.  
 
pK$_{\rm a}$ has been reported for molecules in liquid water using
the AIMD technique.\cite{sprik,klein,parrin1}
It is related to the standard state deprotonation free
energy $\Delta G^{(0)}$ via $-\log_{10}$ $\exp(-\beta \Delta G^{(0)})$,
where $\beta$ is $1/k_{\rm B}T$ and 
\begin{eqnarray} 
\Delta G^{(0)} = -k_{\rm B}T {\rm ln} \bigg\{ C_0
	\int_0^{R_{\rm cut}} dR \, A(R) \,
	\exp[-\beta \Delta W(R)] \bigg\} \, . \label{eq1}
\end{eqnarray}
Here $C_0$ denotes 1.0~M concentration, $R$ is the reaction coordinate,
$A(R)$ is a phase space factor to be discussed below, $R_{\rm cut}$ is
the cutoff distance delimiting the reaction and product valleys in the
free energy landscape, and $W(R)$ is the potential of mean
force which provides the information needed to compute the free energy of
deprotonation.  Regardless of the reaction coordinate used,\cite{sprik,klein}
$W(R)$ generally do not exhibit turning points in the deprotonated
region, and $R_{\rm cut}$ can be taken as the onset of the plateau where
$W(R) \rightarrow 0$.

\begin{figure}
\centerline{\epsfxsize=3.25in \epsfbox{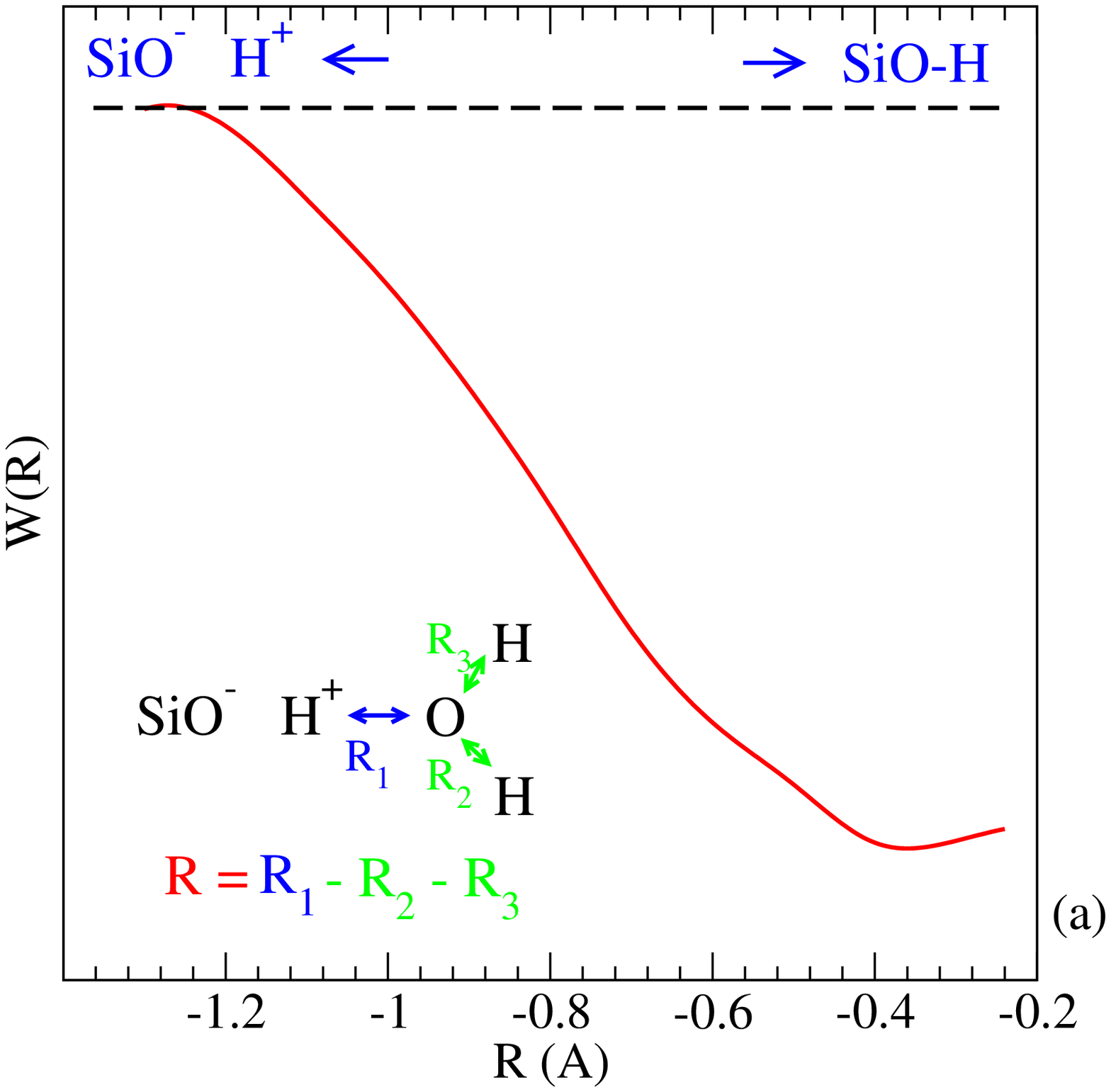}}
\centerline{\hbox{\epsfxsize=0.90in \epsfbox{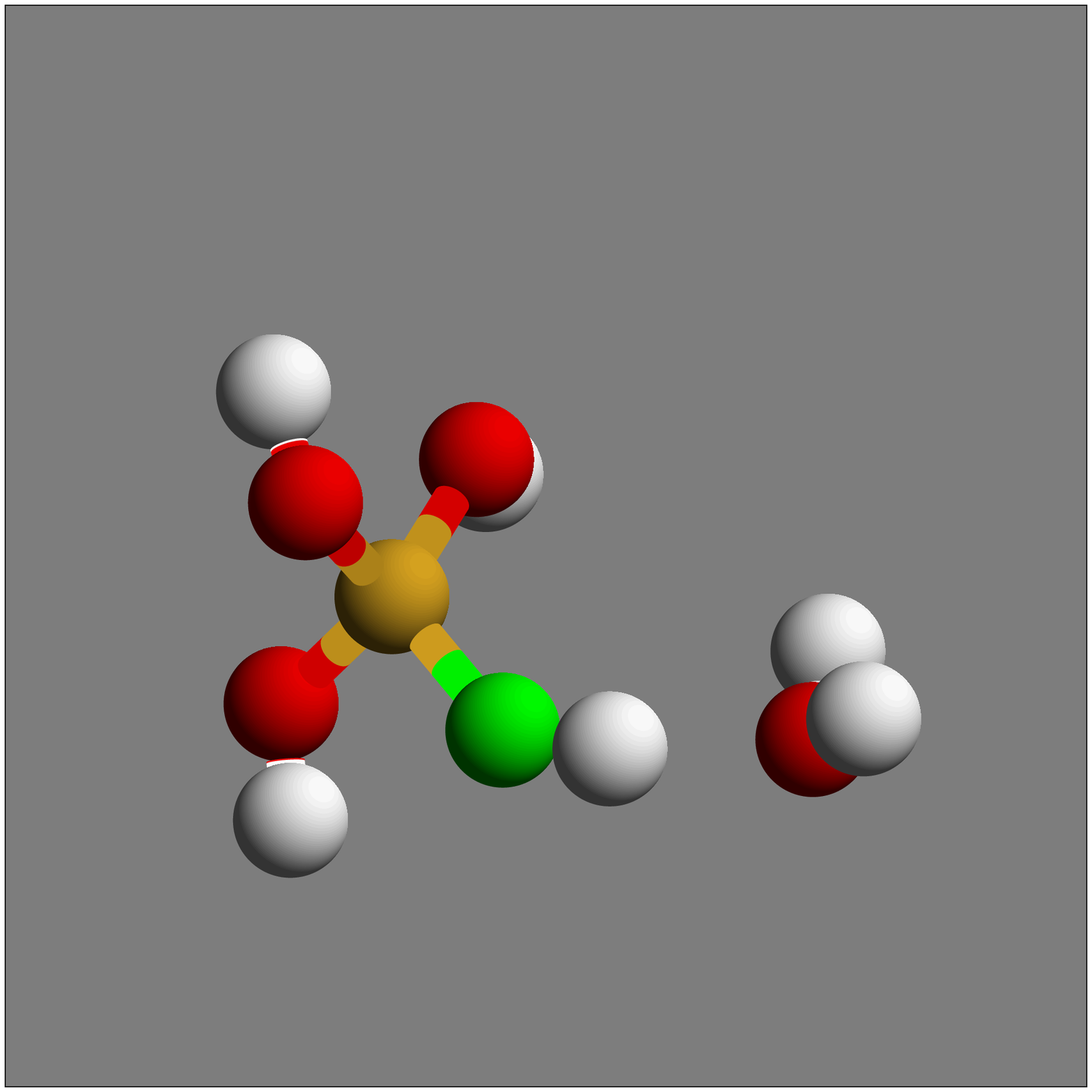}
                  \epsfxsize=0.90in \epsfbox{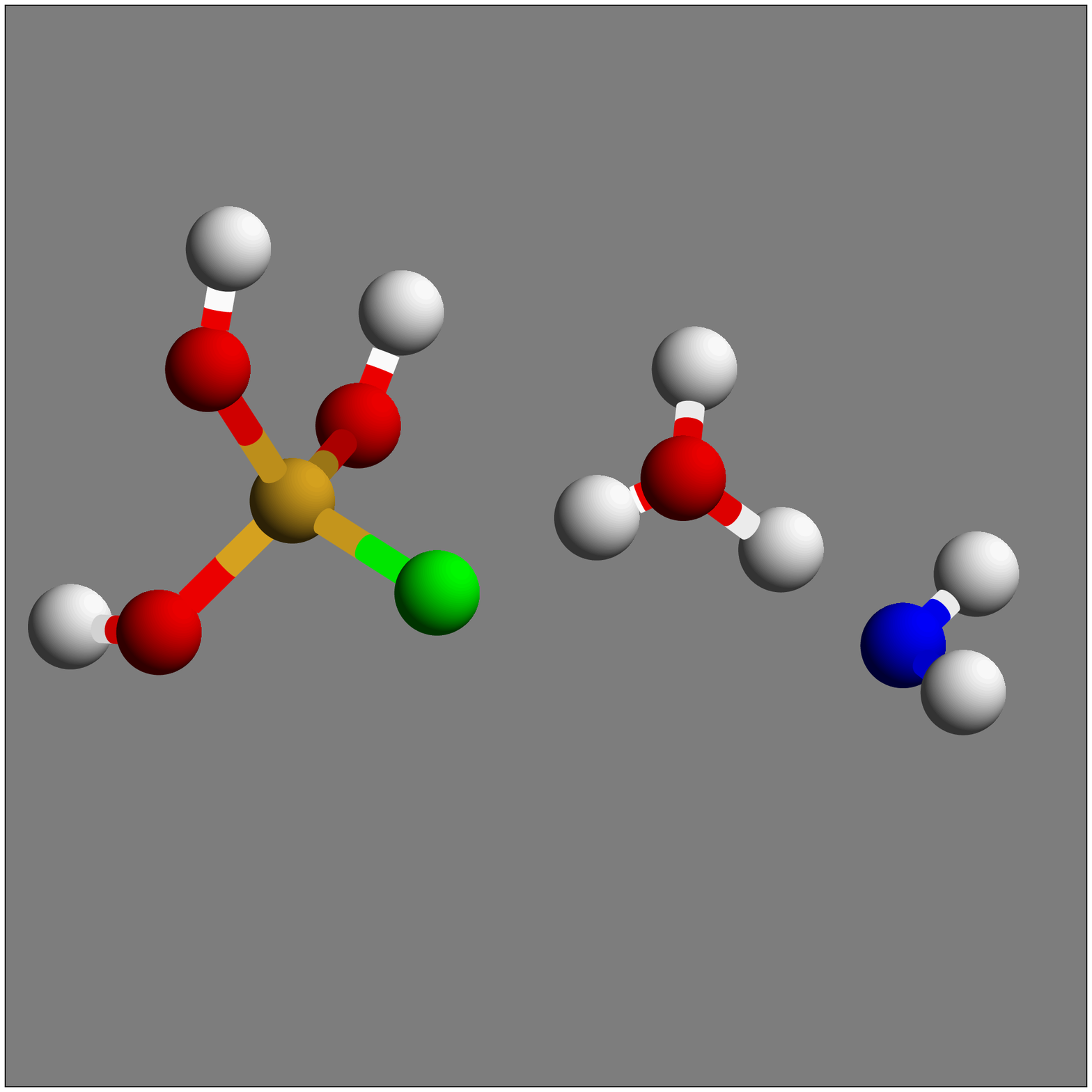}
                  \epsfxsize=0.90in \epsfbox{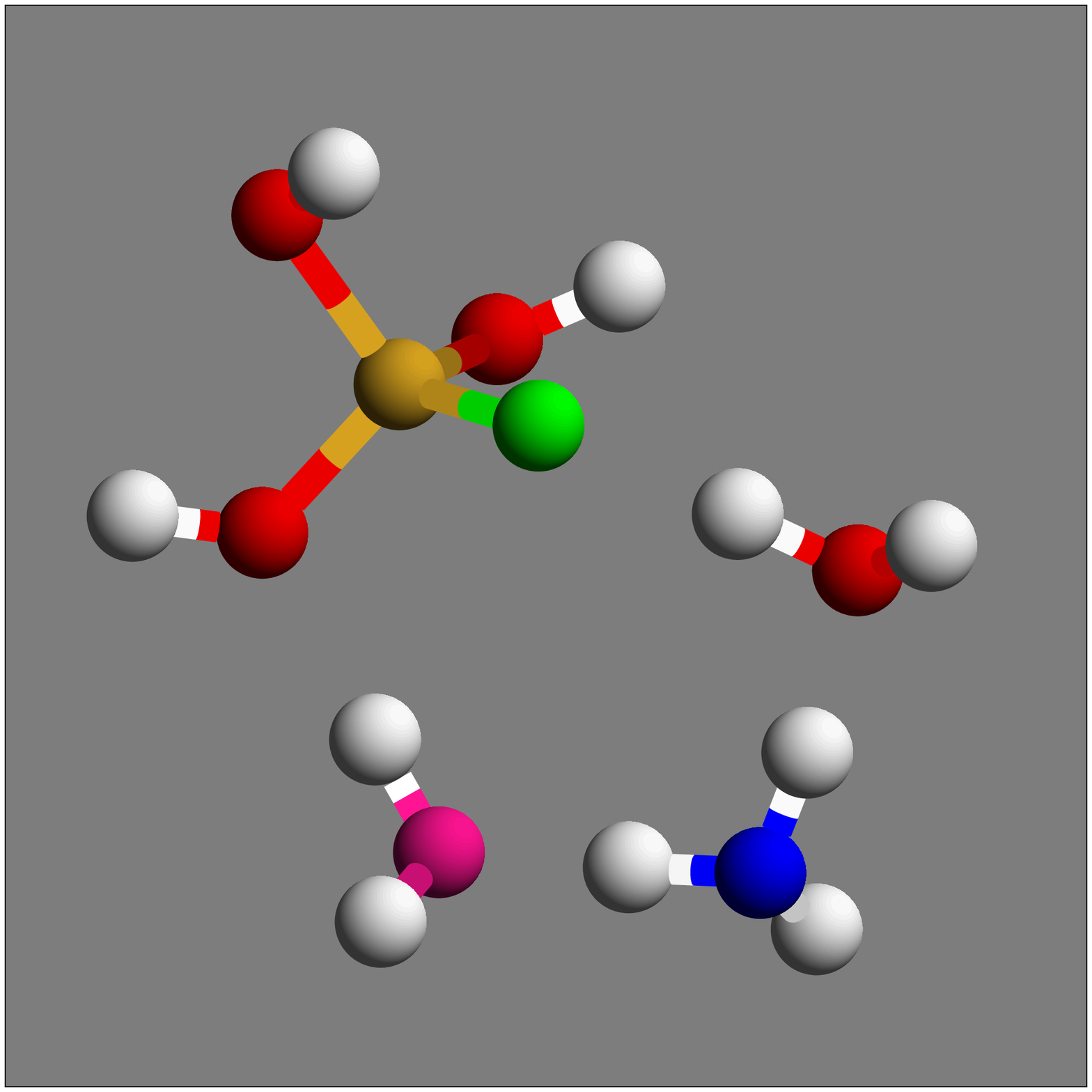} }}
\centerline{\hbox{ (a) \hspace*{1.35in} (b) \hspace*{1.35in} (c)}}
\caption[]
{\label{fig3} \noindent
(a) The 4-atom reaction coordinate $R$, illustrated for silicic acid
in water but is similar for all silanol containing species.  Panels
(b)-(d) are snapshots from AIMD deprotonation simulations,
with outershell H$_2$O molecules removed for clarity reasons.
 As deprotonation proceeds, $R$ progresses from intact SiO-H
($R \sim -0.4$~\AA, panel (b)) to SiO$^-$ H$_3$O$^+$ contact ion pair
($R \sim -1.0$~\AA,
panel (c)) and then, via a Grotthuss proton transfer, to a solvent-separated
SiO$^-$/H$_3$O$^+$ pair ($R \sim -1.32$~\AA, panel (d)).  Yellow, red,
white, and green spheres represent Si, O, H, and the ``O$^-$'' atoms,
respectively.  Additionally, the water O atoms which are second and
third nearest neighbor to the SiO$^-$ oxygen are colored blue and
pink, respectively.  pK$_{\rm a}$ predictions for silicic acid will be
reported in a future publication.
}
\end{figure}

The umbrella sampling method\cite{book1} is used to compute the
$W(R)$ associated with SiOH deprotonation.
A four-atom reaction coordinate $R$ (Fig.~\ref{fig3}a) is found to
work best under our simulation conditions.  It controls what we call
the ``wandering proton'' problem.  We label
the first, second and third neighbor H$_2$O molecules of the SiO$^-$
oxygen (green sphere) shown in Figs.~\ref{fig3}b-d ``water 1''
(O depicted red), ``2'' (blue), and ``3'' (pink).  When $R \sim
-0.4$~\AA\ (Fig.~\ref{fig3}b), the SiOH bond is intact.  As $R$
decreases to $\sim -1.1$~\AA\ (Fig.~\ref{fig3}c), the SiOH proton is
transferred to a ``water 1'' which has been hydrogen-bonded to the SiOH
group, yielding a SiO$^-$-H$_3$O$^+$ contact ion pair.  As $R$ further
decreases (Fig.~\ref{fig3}d), a proton originally residing on
``water 1'' is now transferred to a second water molecule (``water 2''),
creating a water-separated SiO$^-$/H$_3$O$^+$ pair, at which point
the deprotonation reaction is almost complete.  This analysis
appears consistent with insights from a transition path sampling
AIMD simulation,\cite{chandler} as follows.  The Fig.~\ref{fig3}d
configuration, with the excess proton and the SiO$^-$
separated by two hydrogen bonds, form a possible free
energy dividing surface between the intact and the deprotonated
acid species, provided that the electric polarization between the
two states, arising from the surrounding water molecules,\cite{chandler}
has sufficient time to equilibrate.  Our simulation conditions
allow such equilibration, and the excess proton is indeed observed
to diffuse away if $R$ fluctuates to regions significantly more
negative than $\sim -1.4$~\AA. 
Umbrella potentials of the type $(A/2) (R-R_o)^2$ are used to
sample the reaction coordinate $R$.  Just as significantly,
they ensure that  $R>-1.4$~\AA\, and control the extent of
proton transfer from ``water 1'' to all possible ``water 2,'' so that
at most a water-separated ion-pair is obtained.  Otherwise, if the excess
proton is several hydrogen bonds removed from the SiO$^-$ (say if it spends a
significant amount of time on ``water 3'' via proton transfer from
``water 2''), it can start to diffuse (``wander'') through the simulation
cell via the Grotthuss mechanism at O(1)~ps time scale per proton transfer.
With tens of H$_2$O molecules in the simulation cell and 10-20~ps
trajectories, once the excess H$^+$ leaves the second hydration shell
of the SiO$^-$ it does not return, and
equilibrium sampling is not achieved.  This ``wandering'' likely
arises because the higher temperature and longer umbrella sampling
trajectories than are generally used in the AIMD literature
facilitate diffusion of the excess proton away from the SiO$^-$.
Other deprotonation reaction coordinates used in the literature
are discussed in the SI (Sec.~S2).  They are found to give rise to
wandering excess protons under our simulation conditions.
As long as equilibrium sampling is achieved, the deprotonation
free energy cost should not depend on the choice of coordinate.

To apply Eq.~\ref{eq1}, we use a method similar to Ref.~\onlinecite{sprik}:
finding the most probable optimal O$_{\rm (water~1)}$-H$^+$ hydrogen bond
distance $r_{\rm O-H}$ at each $R$, thus locally converting $W(R)$ to
$\bar{W}(r_{\rm O-H})$; performing a spline fit to the resulting
$\bar{W}(r_{\rm O-H})$; and integrating over $r_{\rm O-H}$ with a
$4\pi r^2_{\rm O-H}$ volume element, which takes the place of the phase
space factor $A(R)$ in Eq.~\ref{eq1}.  Equation~\ref{eq1} assumes that
the entropic factors such as rotations of the reactant and products
about the O-H axis are adequately sampled in the AIMD trajectories;
otherwise additional constraints and entropic factors are
introduced.\cite{klein1,co2} SI Sec.~S2 shows that such
constraints do not affect water autoionization free energies, partly
because the variation in $r_{\rm O-H}$ needed to complete the deprotonation
reaction is relatively small.  Furthermore, we always reference predicted
silanol pK$_{\rm a}$ to that of water autoionization\cite{klein} computed
using the same reaction coordinate and elevated temperature.  This minimizes
systematic error arising from the simulation protocol (SI Sec.~S2),
and phase space contributions approximately cancel out.

The metadynamics technique,\cite{meta1,meta2,meta3} a promising and
powerful alternative to umbrella sampling, has been applied to calculate
dissociation free energies on surfaces\cite{marx} and acid-base reactions
of small molecules.\cite{parrin1}  This method, not yet implemented in VASP,
can potentially be used for efficient comparative study of pK$_{\rm a}$
on other material surfaces after it has been adapted to deal with the
wandering proton problem.

Gas-phase, high-level {\it ab initio} calculations are performed to check
and correct chemical bonding energies predicted with the PBE functional
used in the AIMD simulations.  The Gaussian03 program suite is
applied.\cite{g03} The pertinent sample chemical reaction is
\begin{equation}
\mathrm{Si(OH)_4 + H_2O \cdots OH^-} \rightarrow
\mathrm{Si(OH)_3O^- \cdots H_2O + H_2O} \label{rxn1} \\
\end{equation}
Geometries are optimized, and harmonic vibrational frequencies
computed, with density functional theory using the
B3LYP method\cite{lyp,b3lyp} and the 6-311++G($d,p$) basis
set.  At the B3LYP geometries, energies are computed with
the coupled-cluster singles and doubles method including a perturbative
correction for triple substitutions, CCSD(T),\cite{CCSD(T)} using
the aug-cc-pVDZ basis set.  Basis set
incompleteness corrections are added to the CCSD(T) energies. 
Finally, zero-point vibrational energy corrections
computed from the B3LYP/6-311++G($d,p$) frequencies are added.
Using this protocol, the high-level {\it ab initio} calculation
yields a reaction energy of -30.51 for Eq.~\ref{rxn1}.  Gas-phase
VASP-based PBE calculations, conducted with an energy cutoff identical
to that in AIMD simulations, predict -27.17~kcal/mol.  
These numbers do not include the 2.24 kcal/mol zero point energy
(ZPE) corrections. Thus, the overall AIMD reaction energies should be
corrected by a modest -1.10~kcal/mol.  The basis set extrapolation procedure 
may have a systematic error larger than 1 pH unit (SI Sec.~S3) but this does
not affect the relative pK$_{\rm a}$ of different SiOH groups.
 
See the SI for details about constraints introduced to prevent
proton attacks on SiO$^-$, the AIMD intialization protocol, further
justifications for adopting the four-atom reaction coordinate, 
extrapolating quantum chemistry results to the infinite basis set limit.
 
\section{Results}
%\subsection*{$\beta$-cristobalite (100): chemically homogeneous SiOH}
 
{\it Hydroxylated $\beta$-cristobalite (100): chemically homogeneous SiOH.}
We first show that our simulation protocol predicts reproducible
pK$_{\rm a}$ for chemically equivalent SiOH groups on the hydroxylated
$\beta$-cristobalite (100) surface (Fig.~\ref{fig2}a).  This well-studied
model crystalline surface exhibits $\sigma_{\rm SiOH}$=8~nm$^{-2}$, larger
than the experimental value of 4.6~nm$^{-2}$ for amorphous silica.\cite{iler}
At zero-temperature, it features two types
of Q$^2$ silanol groups: alternating hydrogen bond donors and acceptors
arranged in chains\cite{meng} (Fig.~\ref{fig2}a) not found in small
molecules.\cite{sprik,klein,parrin1} This feature is dynamically
preserved in our finite-temperature, aqueous-phase simulations (SI Sec.~S4).  

\begin{figure}
\centerline{\hbox{ \epsfxsize=3.50in \epsfbox{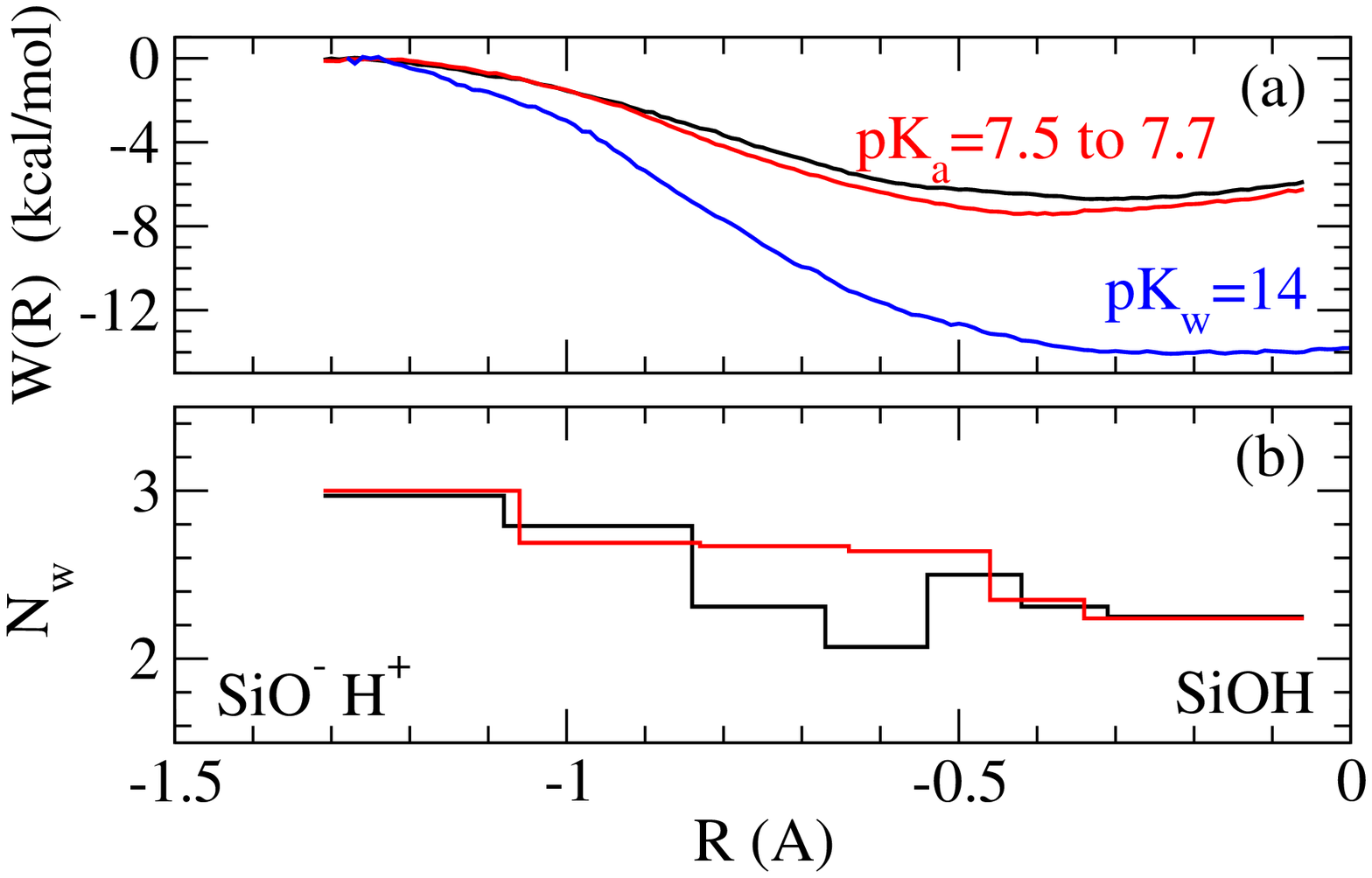} }}
\centerline{\hbox{ (c) \epsfxsize=1.50in \epsfbox{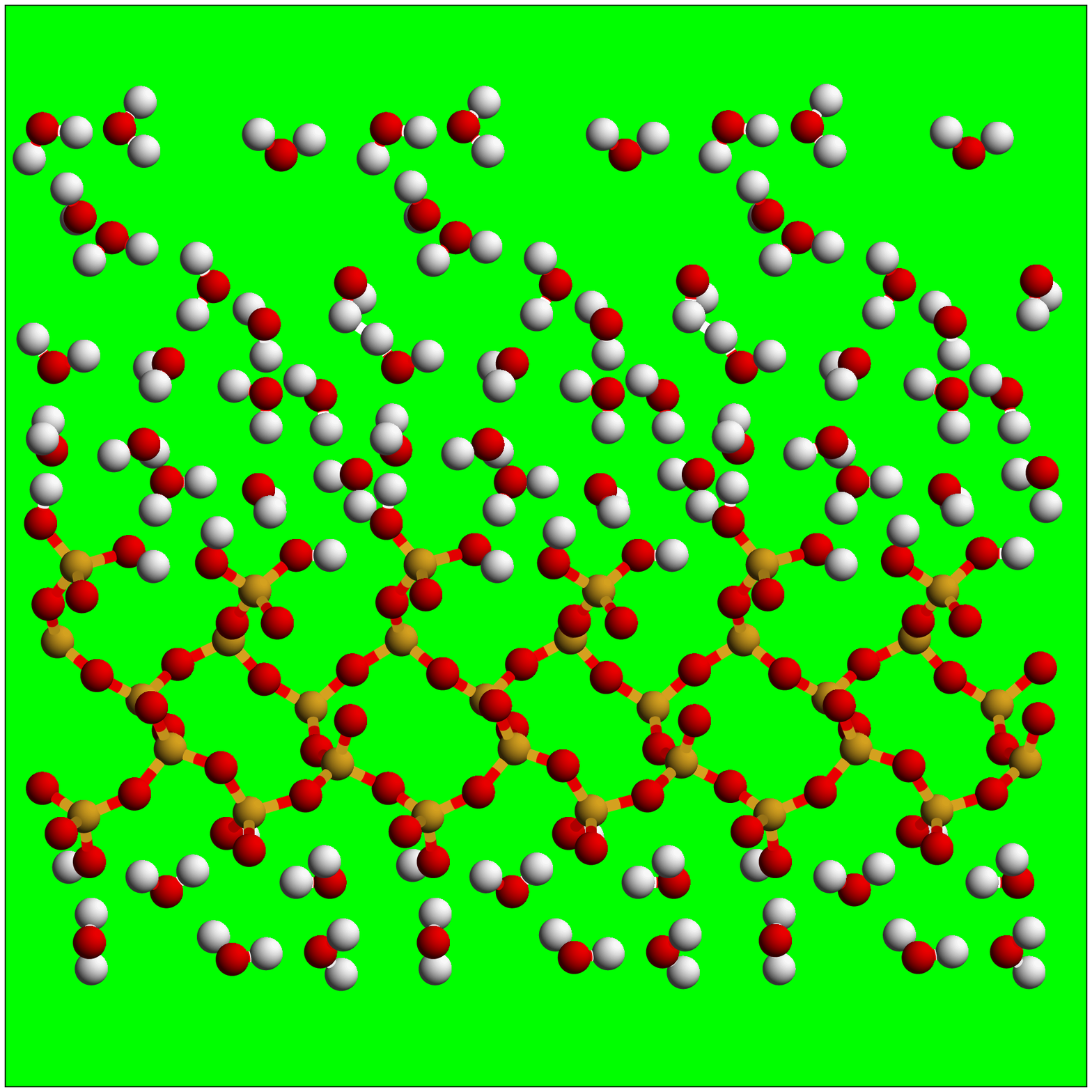} }
		       \epsfxsize=1.50in \epsfbox{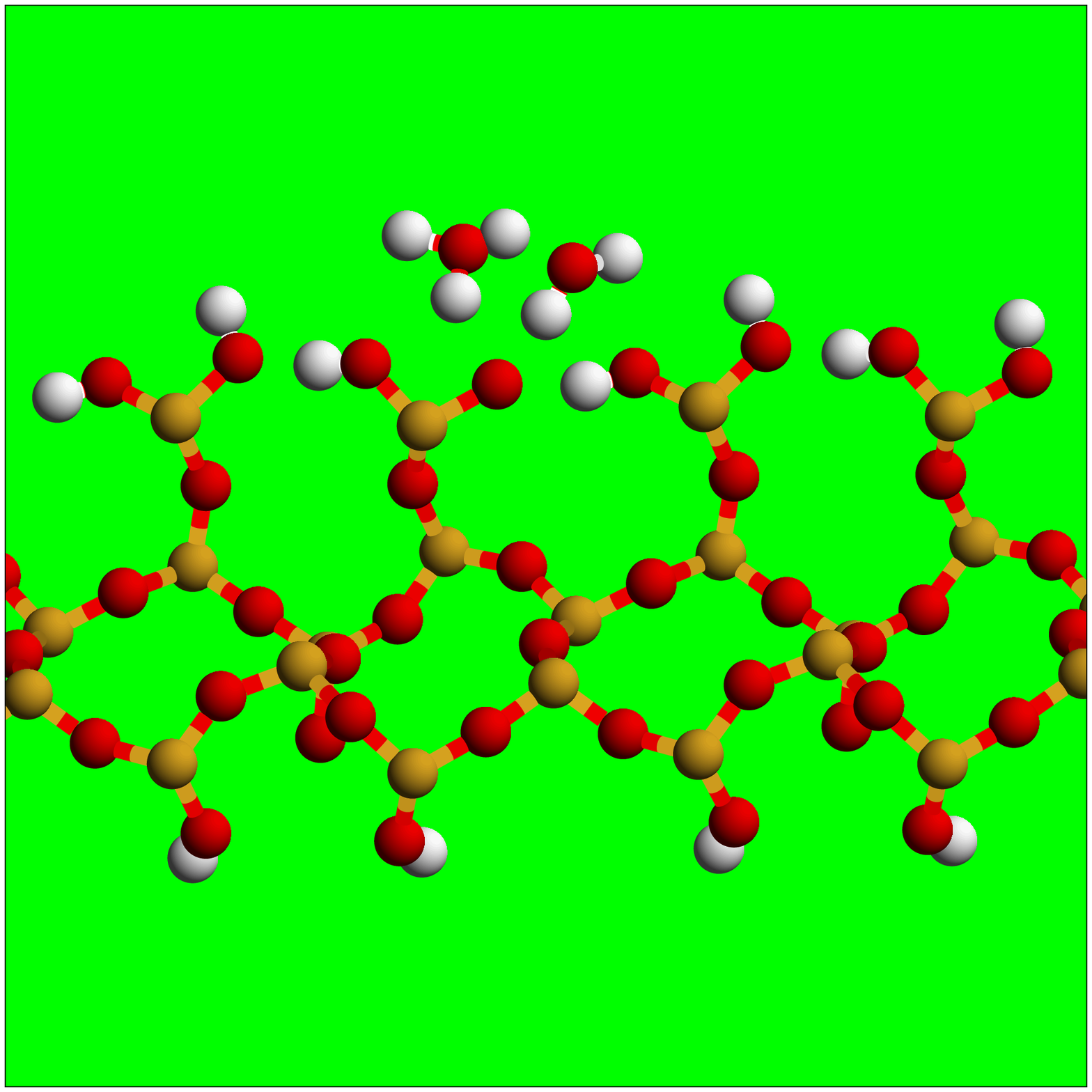} (d) }
\caption[]
{\label{fig4} \noindent
Deprotonation free energy on hydroxylated $\beta$-cristobalite (100)
(Fig.~\ref{fig2}a).  (a) Potential of mean force $W(R)$.
The most negative $R$ values represent SiO$^-$ while
$R >\sim$~$-$0.8~\AA\, indicates intact SiO-H bonds.  
Red and black lines: two different but chemically
equivalent SiOH groups; blue: water autoionization.
(b) Mean hydration number, $N_w$, in each umbrella sampling
window, defined as the number of protons within 2.5~\AA\,
(a typical hydrogen bond distance) of the SiO$^-$ oxygen.
(c) Snapshot of the interface between water and hydroxylated
$\beta$-cristobalite (100), replicated three times in the lateral
direction.  (d) Side view of slab model with silanol groups
forming hydrogen bond chains, highlighting a Q$^2$, H-bonded
SiO$^-$ group in its representative hydrogen bonding environment.
%For clarity, most H$_2$O molecules are omitted.
Yellow, red, and white spheres depict Si, O, and H atoms, respectively.
}
\end{figure}

Figure~\ref{fig4}a shows that two chemically equivalent, hydrogen bond-accepting
SiOH groups on this surface are predicted to exhibit deprotonation $W(R)$
within 0.5~kcal/mol of each other.   At large negative values of
the reaction coordinate $R$, the deprotonated
SiO$^-$ is stabilized with three hydrogen bonds (i.e., $N_w$=3,
Figs.~\ref{fig4}b and~\ref{fig4}c).
At $R >\sim$ $-$0.8~\AA, the SiO-H bond is only partially broken, $N_w <$~3,
%because less electrostatic force exists for hydrogen bonding, 
and the local $W(R)$ is not sensitive to the slight difference in $N_w$ in
the two simulations that arises from statistical noise.  
These observations appear
consistent with a recent two-dimensional potential of mean
force analysis of formic acid deprotonation.\cite{parrin1}  
Accounting for zero-point energy, correcting the
AIMD functional with more accurate quantum chemistry methods,
and referencing Eq.~\ref{eq1} to the water autodissociation
constant pK$_{\rm w}$=14,\cite{klein} we
estimate pK$_{\rm a}$ values of 7.5 and 7.7, close to the
less acidic pK$_{\rm a}$ value reported by Ong {\it et al.}\cite{ong}  
The standard deviation is 0.3~pH unit.  Multiple
deprotonation on this surface is discussed in SI Sec.~S5.
 
%\subsection*{Heterogeneity: Isolated, Vicinal, Q$^2$, and Q$^3$ SiOH
%all exhibit pK$_{\rm a} >$ 7}
 
\begin{figure}
\centerline{\epsfxsize=3.50in \epsfbox{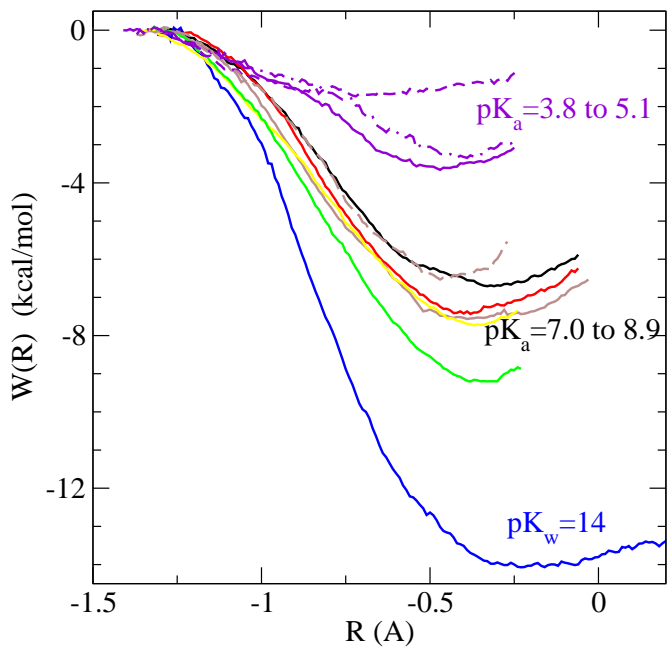}}
\caption[]
{\label{fig5} \noindent
Distribution of pK$_{\rm a}$.  Black and red lines: $W(R)$ for SiOH on the
hydroxylated $\beta$-cristobalite (100) surface (Fig.~\ref{fig2}a); 
green: same surface but with hydrogen bond-donating SiOH artificially
removed (Fig.~\ref{fig2}b); brown, solid and dashed: reconstructed
$\beta$-cristobalite (100) (Fig~\ref{fig2}c), with 6 and 4 layers of
water in the simulation cell, respectively; yellow: (H$_3$SiO)$_3$SiOH
(Fig.~\ref{fig2}e); violet: reconstructed quartz (0001),
(Fig.~\ref{fig2}f), with the SiOH residing on a silica trimer ring (solid)
or otherwise (dashed and dot-dashed).  Blue: $W(R)$ for water autoionization.
}
\end{figure}

{\it Heterogeneity: Isolated, H-bonded, Q$^2$, and Q$^3$ SiOH
all exhibit pK$_{\rm a} >$ 7.}  We next show that, contrary to previous
hypotheses,\cite{ong,lorenz,fan1,mori1,nanosilica,others,shaw2,rosenholm} 
isolated,  H-bonded, Q$^2$, and Q$^3$ silanol groups all exhibit 
pK$_{\rm a} >$ 7.0.  We first create an isolated silanol group by
replacing a hydrogen bond-donating SiOH group on the hydroxylated
$\beta$-cristobalite (100) surface with a SiH so that its
neighboring SiOH group is no longer H-bonded (Fig.~\ref{fig2}b).
Figure~\ref{fig5} shows that this isolated SiOH exhibits
pK$_{\rm a}$=8.9$\pm 0.3$, and is {\itshape less} acidic by 1.2-1.4 pH unit
than when the SiOH hydrogen
donor is present (Fig.~\ref{fig2}a).  This is entropically reasonable because
a hydrogen bond-donating SiOH partner stabilizes the neighboring SiO$^-$
alongside two water molecules, while three water molecules are required for an 
isolated SiO$^-$.  AIMD correctly accounts for this effect because it
models H$_2$O and SiOH on the same dynamical footing and because water-water
and water-silanol hydrogen bond energies are similar.\cite{meng}
Hydroxyls on oxides with more ionic character than SiO$_2$ form stronger
hydrogen bonds, and indeed the relative abundance of inter-hydroxyl hydrogen
bonding may partially be responsible for the crystal facet-dependent acidity of
$\alpha$-Al$_2$O$_3$.\cite{fitts}  This will be the subject
of future, comparative studies.
 
To compare Q$^2$ and Q$^3$ silanol groups, we reconstruct the hydroxylated
$\beta$-cristobalite (100) surface by condensing every other pair of
H-bonded SiOH groups into a SiOH and a H$_2$O molecule (Fig.~\ref{fig2}c).
This involves the elimination of a hydrogen-bond donating OH group plus the
proton on its adjacent, hydrogen-bond accepting SiOH (Fig.~\ref{fig2}d).
The resulting undercoordinated Si and O atoms are joined together
to form a covalent bond, in the process pulling apart the remaining H-bonded
SiOH pairs so they are now isolated from one another.  
A similar structural motif has been considered in the literature.\cite{chuang}  
This surface has $\sigma_{\rm SiOH}$=4~nm$^{-2}$, with all SiOH groups
being Q$^3$, isolated, and residing on silica rings containing at
least 5~Si atoms.  Such rings should be unstrained, unlike 3-member
(Si-O)$_3$ rings discussed below.  The pK$_{\rm a}$ is found to be
8.1$\pm 0.5$ (Fig.~\ref{fig5}).  

We also consider a (SiH$_3$)$_3$SiOH molecule featuring
an isolated, Q$^3$ SiOH group (Fig.~\ref{fig2}e), which exhibits a comparable
pK$_{\rm a}$=7.9$\pm 0.5$.
Thus, in general, Q$^3$ and Q$^2$ silanol groups do not exhibit pK$_{\rm a}$'s
that differ by 4 pH units as previously
proposed.\cite{lorenz,fan1,mori1,nanosilica,others,shaw2,rosenholm}
Over the range 4~nm$^{-2}$ $\leq \sigma_{\rm SiOH} \leq$ 8~nm$^{-2}$,
the precise value of $\sigma_{\rm SiOH}$ has little effect on pK$_{\rm a}$.
 
To some extent, all our crystalline silica models are nano-slits with thin
water slabs confined between the surfaces.  As the water content decreases,
the dielectric solvation of SiO$^-$ and H$^+$ species should decrease, while
intact SiOH groups should be weakly affected.  Thus one expects a lower acidity
and a higher pK$_{\rm a}$ in strongly nano-confined aqueous media.\cite{yb}
To examine confinement effects, a pK$_{\rm a}$ calculation is performed for a
smaller reconstructed $\beta$-cristobalite (100) simulation cell, 20~\AA\,
in the $z$-direction, containing 4 layers of water.  This system 
actually yields a lower pK$_{\rm a}$=7.0$\pm$0.4---but is lower only
by 1.1~pH units (dashed brown line in Fig.~\ref{fig5}).  It is possible
the unexpected pK$_{\rm a}$ decrease between the 6- and 4-water layer models
arises from anomalies in the hydrogen bonding network not apparent from
visual inspection of water configurations.  The water density
is also affected by the confinement.\cite{garofalini2}  In the 4-layer model,
all water molecules are at most two layers away from the crystalline silica
surfaces, and AIMD conducted with GCMC-predicted water content is found to
yield a second layer H$_2$O density that is 18\% above 1.0~g/cc.  However,
this is unlikely to change the dielectric response sufficiently to
lower the pK$_{\rm a}$ by 1 unit.  Assuming one can apply the Born
hydration formula for excess proton hydration in this heterogeneous medium,
$\Delta G_{\rm hyd} \sim X (1-1/\epsilon_o)$ where $X\approx -264$~kcal/mol
for the proton.\cite{tiss}  If the 6-layers of water have $\epsilon_o$=80,
the 4-layer system must exhibit $\epsilon_o$=140 to make hydration more
favorable by 1~pH unit when in fact confinement generally reduces the
dielectric constant of water.\cite{pore} In any case, the discrepancy
in pK$_{\rm a}$ is actually within two standard deviations and may simply
arise from statistical uncertainties.  This test suggests that confinement
effects are not large for the slit pore geometry\cite{dekker1}
down to about 1~nm slit widths.  Therefore our reported pK$_{\rm a}$
for 6-water layer systems should be good approximations of pristine
crystalline silica surfaces in contact with bulk liquid water.
We also stress that all slab geometries in Figs.~\ref{fig2}a-c
have been studied using 6-water layer simulation cells, and their
{\it relative} pK$_{\rm a}$ should be mostly free of system size effects.
However, two-dimensional confinement in cylindrical amorphous silica
nanopores\cite{pore,lorenz,schulten,hartnig,dipole} may have a stronger impact
on pK$_{\rm a}$.\cite{yb}

%\subsection*{High acidity and chemical reactions on strained surfaces}
 
{\it High acidity and chemical reactions on strained reconstructed quartz
surfaces.}
Finally, it seems imperative to demonstrate the possibility of an
unusually acidic silanol group.  The following ``computational existence
proof'' is necessarily more speculative than the conclusions
about Q$^2$, Q$^3$, isolated, and H-bonded SiOH pK$_{\rm a}$ discussed
above, but it emphasizes the likely role of defected regions when
accounting for pK$_{\rm a}$=4.5.

Having considered $\sigma_{\rm SiOH}$=8
and~4~nm$^{-2}$, we examine an even lower SiOH surface density.  A recent
experimental study has attached crystal violet dyes to deprotonated
silanol groups.  Based on the flat, 120~nm$^2$ surface
area of the dye molecule, it is proposed that strong acidity is 
correlated with local SiOH surface density
$\sigma_{\rm SiOH}$ $<$ 0.83~nm$^{-2}$.\cite{dong}  In our DFT
calculations, the optimal geometry of crystal violet when covalently
bonded to (HO)$_3$SiO$^-$ is not flat but is substantially distorted.
However, this suggestion of low SiOH surface density being
associated with the more acidic SiOH appears consistent with
other experimental and theoretical observations discussed below.

Since most cuts through crystalline forms of silica yield surfaces with
substantial hydroxylation,\cite{nangia} we investigate a reconstructed,
completely dehydroxylated quartz (0001) model, featuring (Si-O)$_3$
trimer rings, predicted to be metastable in vacuum.\cite{deleeuw1,deleeuw2} 
2.3~SiOH groups per square-nanometer are re-introduced by removing two
surface Si atoms and performing further reconstruction and hydroxylation to
keep all atoms fully coordinated.  This yields two types of silanol
groups which are hydrogen-bonded to each other; one member of the pair
resides on a cyclic trimer while the other does not (Fig.~\ref{fig2}f).
Regions devoid of SiOH and dominated by siloxane (Si-O-Si) bridges are
hydrophobic, and this model surface may therefore be consistent with
hydroxyl groups in hydrophobic pores reported to be unusually acidic on
other oxides.\cite{alumina2}  We conduct one deprotonation umbrella sampling
simulation of a silanol group residing on a cyclic trimer (Fig.~\ref{fig5},
solid violet curve), and two simulations where the SiOH does not
reside on a trimer (dashed and dot-dashed violet). Figure~\ref{fig5} indeed
shows that these SiOH groups exhibit pK$_{\rm a}$ $\sim 5.1$$\pm 0.3$,
4.8$\pm 0.4$, and $3.8$$\pm 0.4$, respectively --- close to the experimental
value of 4.5.  More significantly, their average pK$_{\rm a}$ are
separated from the median of all other SiOH groups previously examined
in this work by 3.4~pH~units.  
 
Unlike cyclic silica tetramers or larger Si-O rings, cyclic silica trimers
are strained.\cite{book,strain,wallace1} At zero temperature, in the absence
of water, the Si atom of the SiOH group residing on a silica trimer ring
exhibit Si-O-Si angles of 137.5$^o$, 130.4$^o$, and 132.0$^o$.
For the SiOH group not residing on a silica trimer, the
angles are 154.9$^o$, 131.8$^o$, and 147.8$^o$.  A few 
of these angles deviate substantially from the ideal, unstrained
Si-O-Si value of approximately 145$^o$.  This likely accounts
for the low pK$_{\rm a}$ computed for these SiOH groups.
Other trimer rings not decorated with SiOH are also strained, and the
slow spatial decay of their surface strain fields\cite{rickman} may
also contribute to the high acidity of SiOH group not residing on them.

\begin{figure}
\centerline{\hbox{ (a) \epsfxsize=1.50in \epsfbox{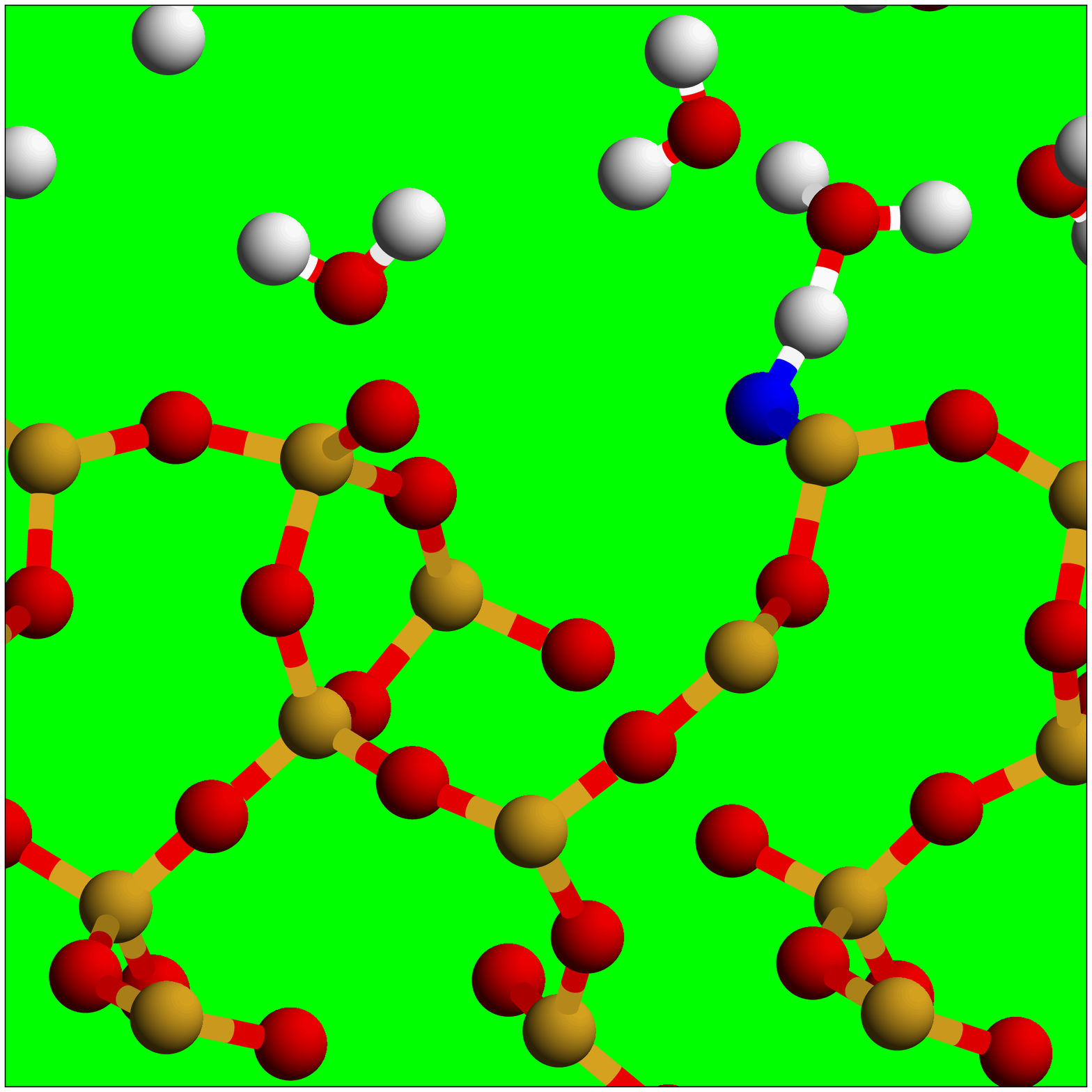}
                   (b) \epsfxsize=1.50in \epsfbox{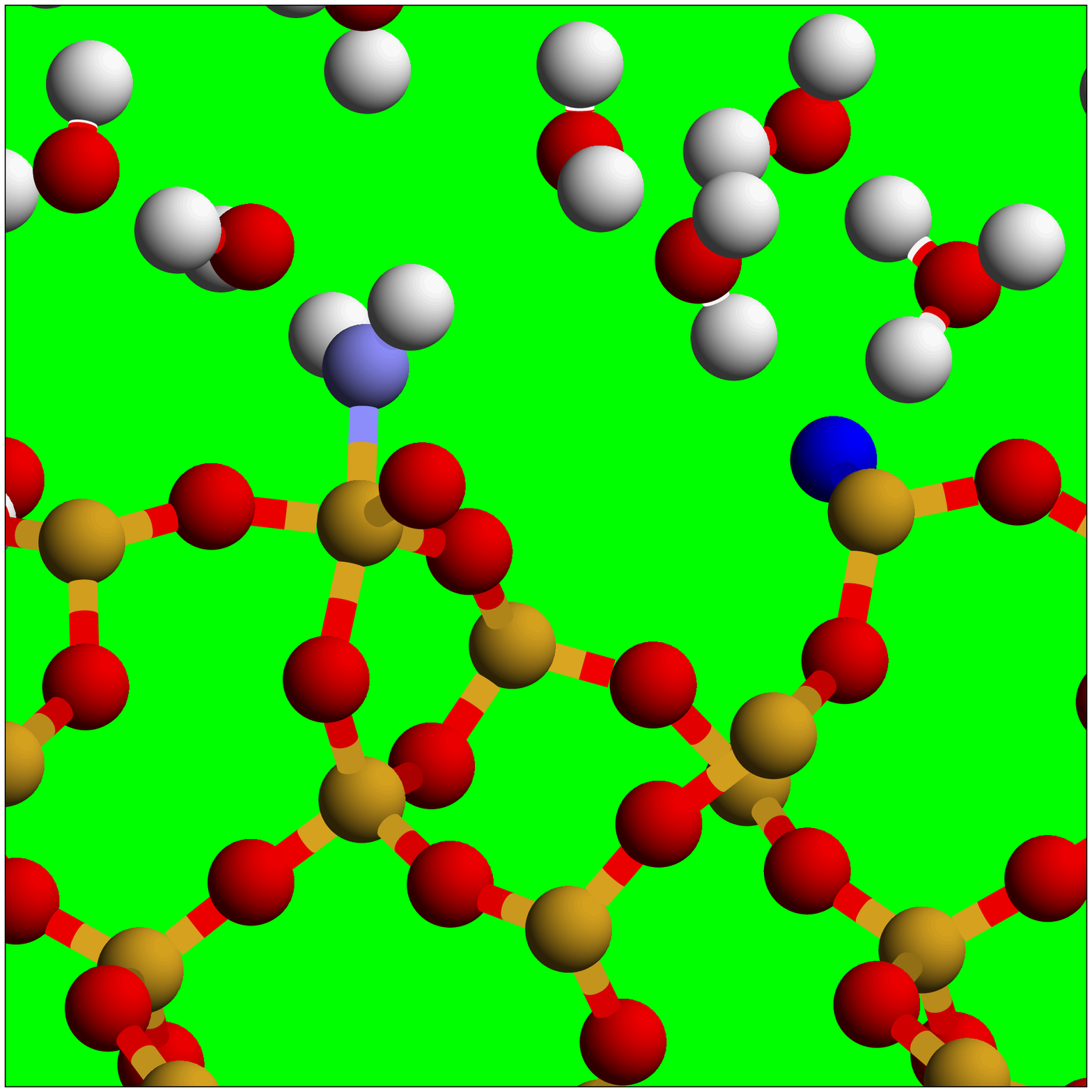} }}
\centerline{\hbox{ (c) \epsfxsize=1.50in \epsfbox{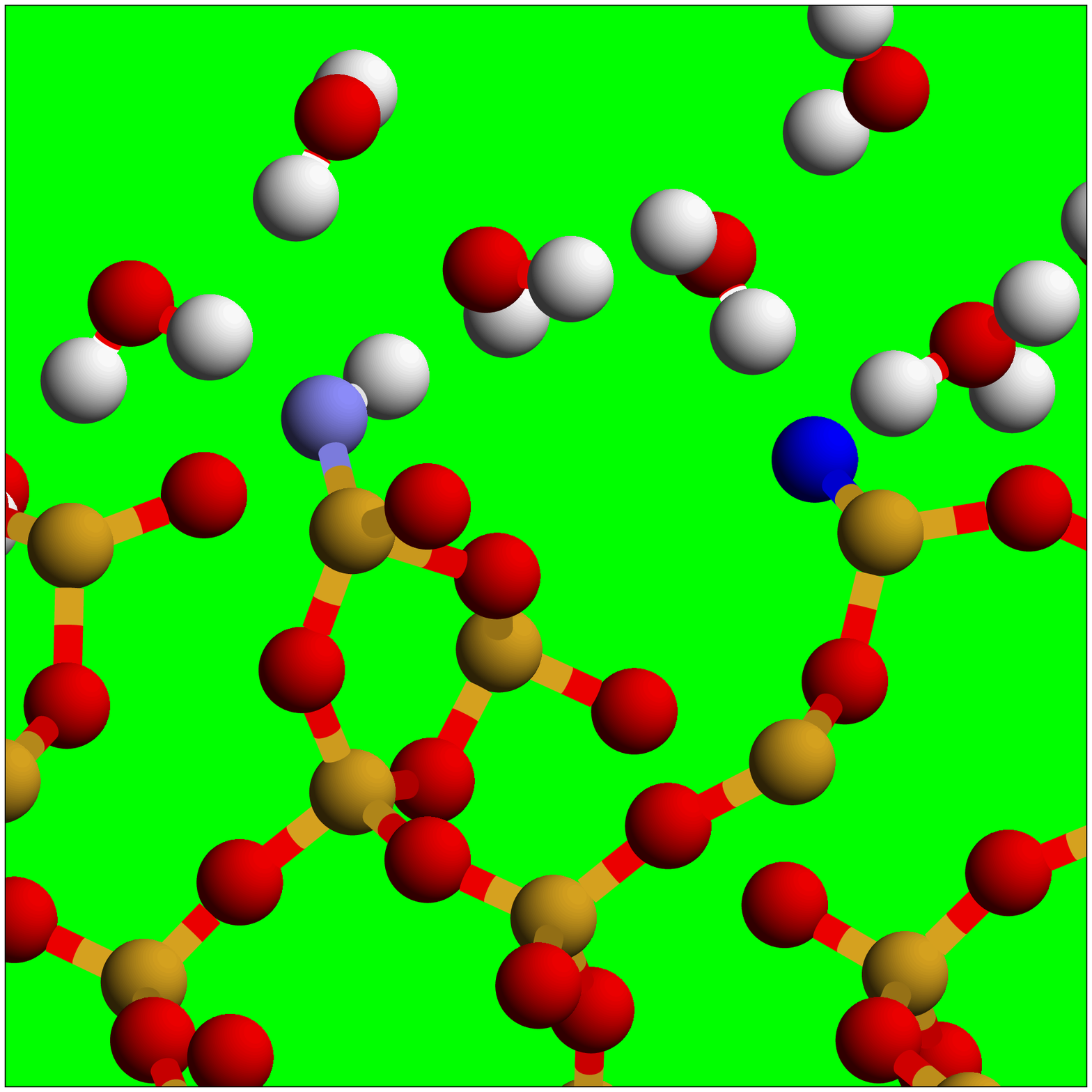}
                   (d) \epsfxsize=1.50in \epsfbox{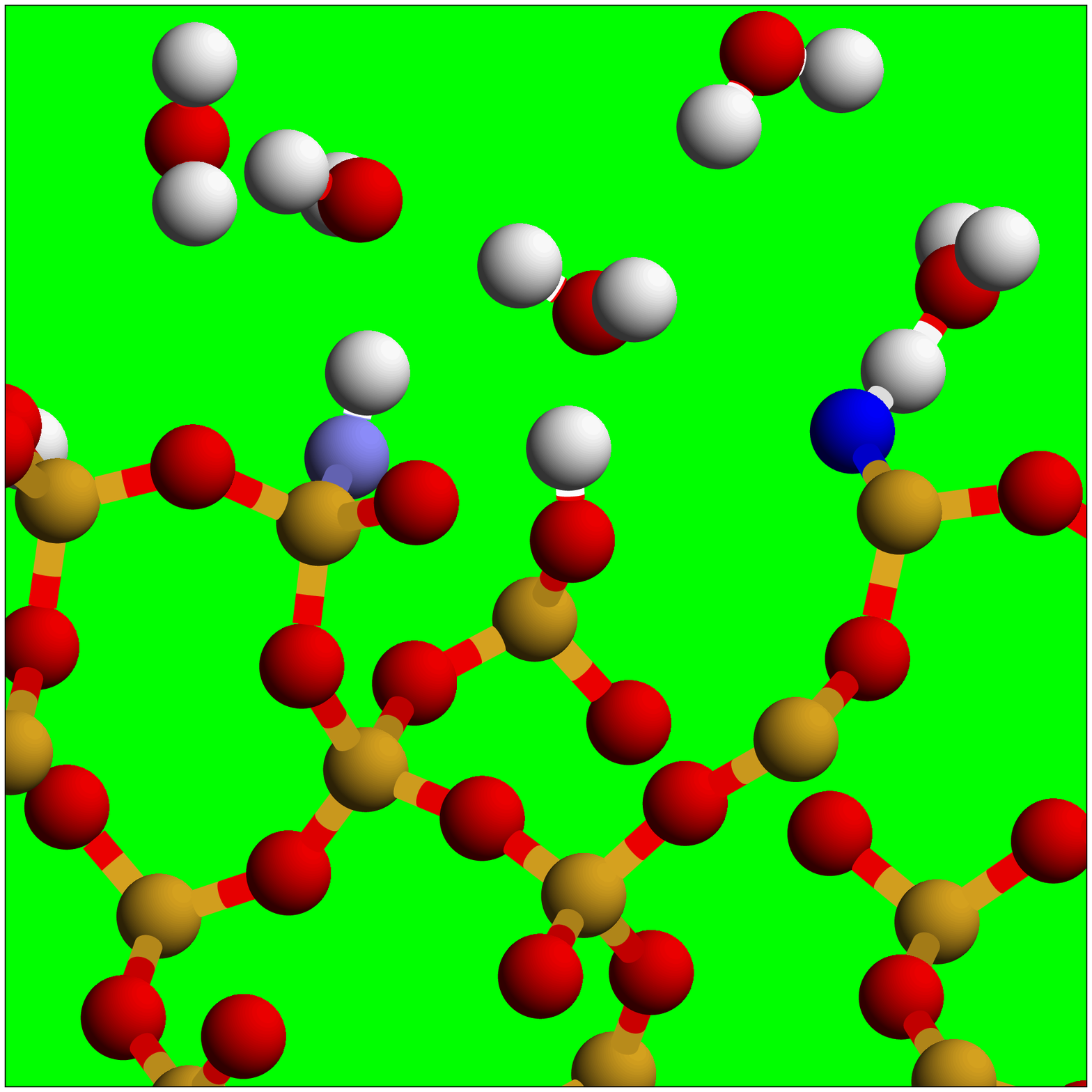} }}
\caption[]
{\label{fig6} \noindent
Water incorporation onto reconstructed quartz.  
(a) Initial equilibrated snapshot of SiO$^-$ H$^+$ contact ion pair
on the reconstructed quartz (0001) surface where the SiO$^-$ resides
on a 3-member ring (truncated in this figure).  The tagged SiO$^-$
oxygen is in deep blue; its H-bond SiOH donor is out of the frame.
All other Si, O, and H atoms are in yellow, red, and white, respectively.
(b) Snapshot after $\sim$ 3~ps.  To the left of the SiO$^-$, a water
molecule (light blue) has attached itself to a Si atom on another
silica trimer ring, which is unhydroxylated. That Si becomes 5-coordinated.
(c) This water molecule loses a proton to bulk water; one of the Si-O
bonds not on the trimer ring becomes stretched.  This is reversible as
long as another SiOH is not created.  (d) In another sampling window for
this SiOH, the trimer ring breaks open instead, forming a new SiOH after
extracting a proton from water.
}
\end{figure}

Within hours in moist air,\cite{book,strain,wallace1} cyclic trimer-containing
surfaces are known to incorporate water and break open to reduce strain
and increase the local $\sigma_{\rm SiOH}$.  At the water-reconstructed
quartz interface, during umbrella sampling deprotonation of the SiOH group
residing on a 3-member ring (Fig.~\ref{fig2}f), we indeed observe a water
molecule forming a transient bond with another Si atom on a Si-O trimer
6~\AA\, away from the tagged SiO$^-$ within picoseconds (Fig.~\ref{fig6}b).
The resulting 5-coordinated Si has been observed in
simulations\cite{deleeuw2,garofalini1,garofalini} and found to be the
intermediate in the trimer ring-opening mechanism on wet silica
amorphous surfaces in reactive force field and molecular orbital
calculations.\cite{garofalini1,garofalini,lasaga} Our AIMD trajectories show
that this mechanism remains operative at explicit liquid water-silica
interfaces; proton hopping via the Grotthuss mechanism occurs readily,
enabling the H$_2$O adsorbed on the surface Si to lose a proton to bulk water,
forming a new SiOH group within $\sim 10$~ps (Fig.~\ref{fig6}c).  Then, in
2 of the 4 sampling windows, a Si-O bond on the now 5-coordinated Si breaks
to open the OH-incorporated (Si-O)$_3$ ring and irreversibly
introduce another new SiOH (Fig.~\ref{fig6}d).  Our trajectories thus
differ from a recent molecular dynamics study of the liquid
water-reconstructed quartz interface, where the adsorbed water molecule,
not described by a reactive force field, ultimately
desorbs from the 5-coordinated surface Si atom without inducing chemical
reactions.\cite{deleeuw2} 
 
\begin{figure}
\centerline{\hbox{ (a) \epsfxsize=1.50in \epsfbox{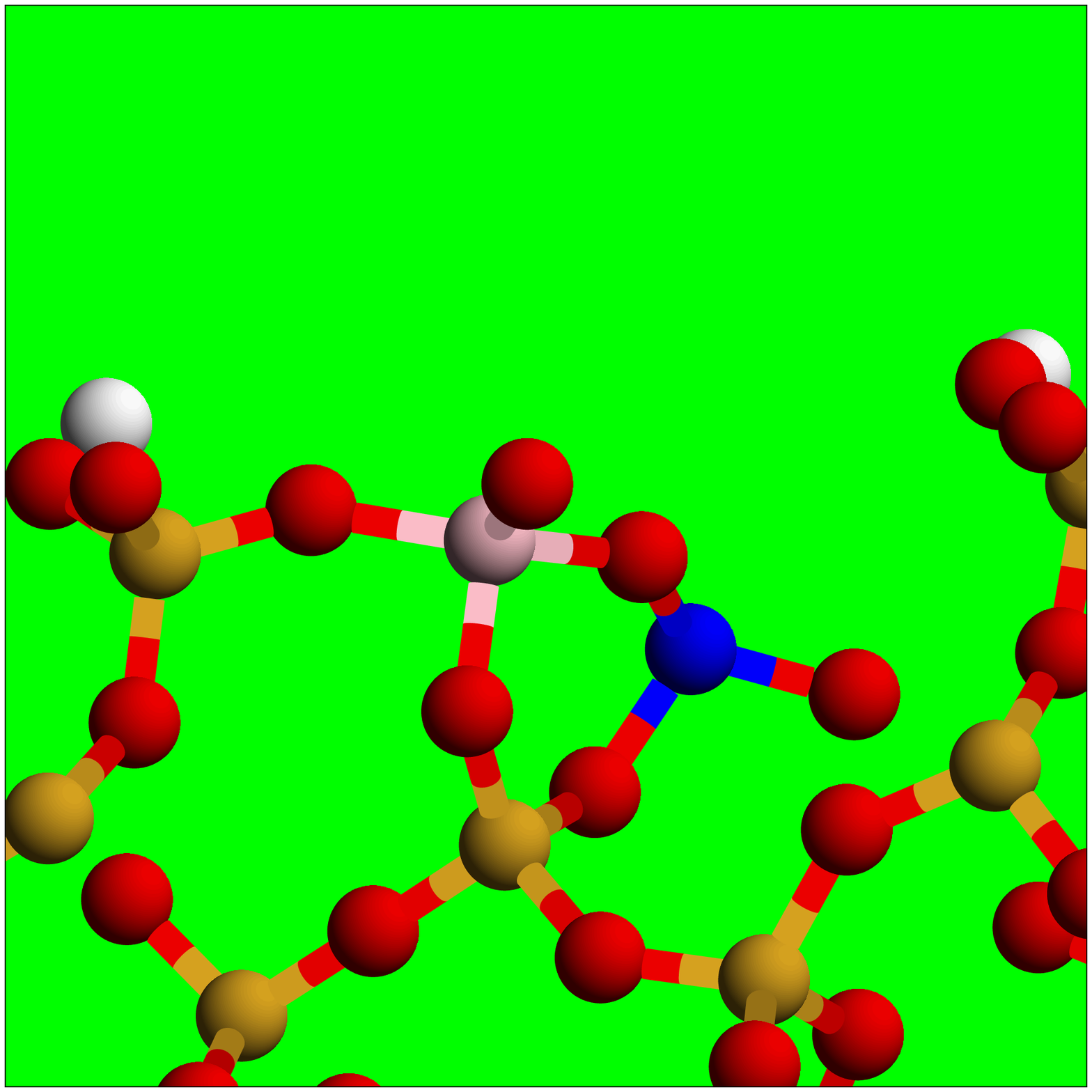}
                   (b) \epsfxsize=1.50in \epsfbox{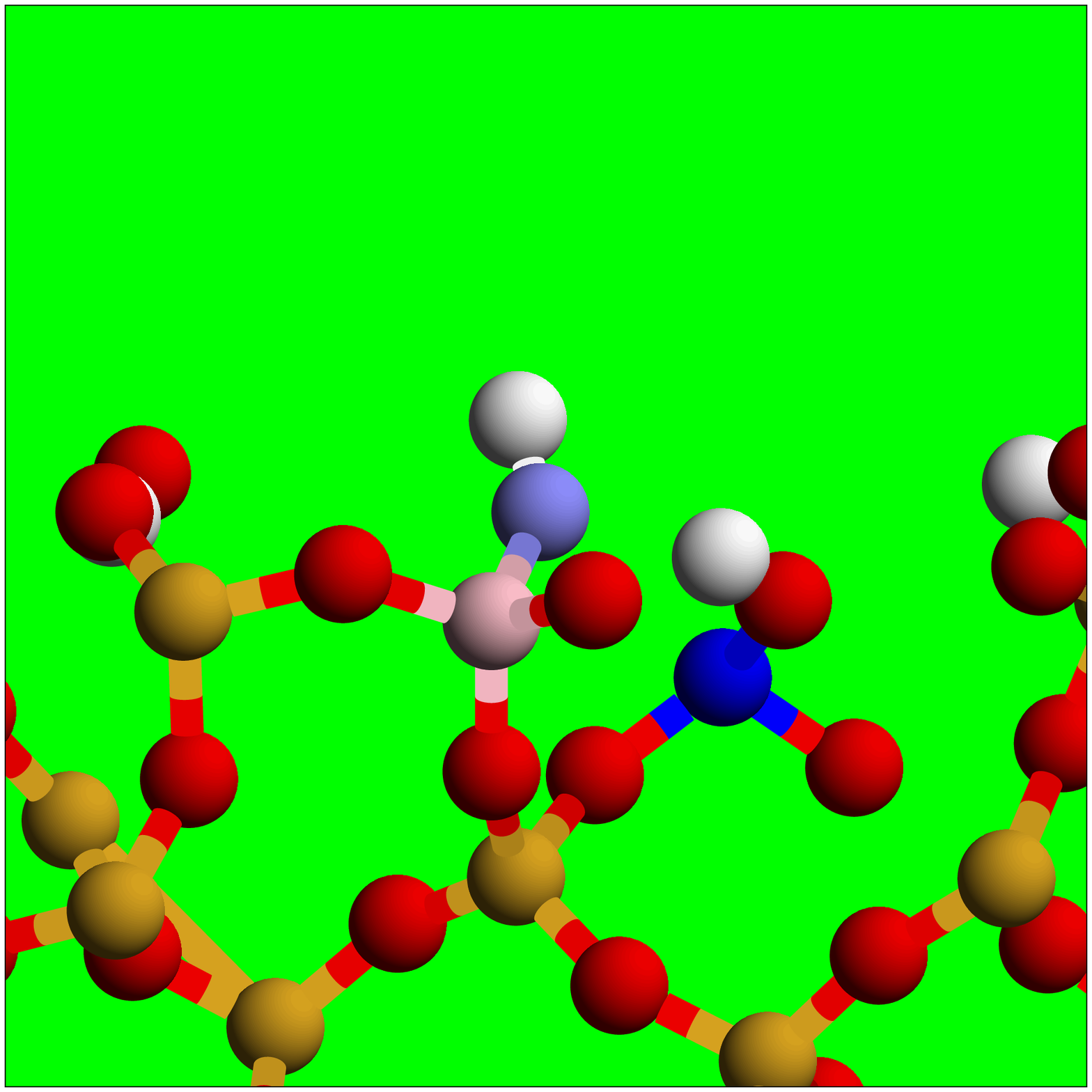} }}
\caption[]
{\label{fig7} \noindent
Trimer silica ring and Si-O-Si angles.  Configurations are taken
from AIMD pK$_{\rm a}$ calculations where the SiOH group involved
in the pK$_{\rm a}$ calculation does not reside on a silica trimer
ring.  That tagged SiOH is off-frame; all H$_2$O molecules are removed
for clarity.  (a) Snapshot along AIMD trajectory where the 3-member ring
remains intact (case A).  The 3 Si-O-Si angles within the ring average to
121.1$^o$ (between Si atoms on the trimer ring colored yellow
and dark blue), 134.0$^o$ (yellow/pink), and 132.2$^o$ (pink/dark blue).
(b) Incorporation of a H$_2$O (its oxygen colored light blue) opens
the ring (case B).  The surviving Si-O-Si angles are 140.5$^o$ (yellow/dark
blue) and 136.0$^o$ (yellow/pink).  O and H atoms are represented by
red and white spheres respectively.
}
\end{figure}

These irreversible side reactions prevent strict equilibrium sampling
needed for $W(R)$ calculations.  Fortunately, for the SiOH residing on a
cyclic trimer (Fig.~\ref{fig6}),
analysis of the pre- and post ring-breaking statistics reveals that the
nearby chemical reaction has little effect on its pK$_{\rm a}$.
We further analyze trimer ring-opening effects on the pK$_{\rm a}$
of another SiOH group, this one not residing on a surface
3-member ring.  A H$_2$O incorporation reaction also
occurs in the neighborhood of this tagged SiOH (Fig.~\ref{fig7}).
We split the sampling windows into two groups: (A) those without irreversible
hydrolysis of a nearby silica ring (Fig.~\ref{fig7}a); and (B) those with
trimer ring breaking and formation of two new SiOH groups (Fig.~\ref{fig7}b).
Then two complete sets of sampling windows spanning the entire deprotonation
pathway are spawned from these seed windows, yielding two pK$_{\rm a}$:
case A, pK$_{\rm a}$=3.8 (Fig.~\ref{fig5}, dashed violet curve); and
case B, pK$_{\rm a}$=4.8 (Fig.~\ref{fig5}, dot-dashed violet).  
The results show that H$_2$O incorporation and a single ring-breaking
event nearby does appear increase the pK$_{\rm a}$ of the tagged SiOH
not residing on a trimer ring. However, the increase is only 1.0~pH unit,
almost within statistical uncertainties.  In case (A), the three Si-O-Si
angles on the silica trimer ring average
to 121.1$^o$, 134.0$^o$, and 132.2$^o$ along the trajectory; the first
refers to the angle where both Si are below the silica surface (``buried'').
In case (B), this angle linking the buried Si atoms relaxes significantly to
140.5$^o$.  The second angle, which involves the surface Si and a buried Si,
remains almost unchanged at 136.0$^o$.  (The third linkage is destroyed during
ring opening.) Si atoms which no longer participate in strained Si-O-Si
linkages should be more stable against H$_2$O attack.

From these considerations,
we conclude that, despite interference from H$_2$O incorporation reactions,
there remains a statistically significant difference in the pK$_{\rm a}$'s
on this surface, and the pK$_{\rm a}$'s ranging from 7.0 to 8.9 for all
other silanol groups investigated before (Fig.~\ref{fig5}).  This finding
suggests that strain, low local silanol surface density, hydrophobicity,
and low pK$_a$ are correlated on amorphous silica surfaces.  Indeed,
atomistic model surfaces with low local $\sigma_{\rm SiOH}$ regions almost
always exhibit 3-member rings.\cite{deleeuw1,garofalini1,singer}
These regions may be in dynamic equilibrium with solvated silica fragments
in solution, constantly being dissolved/hydrolyzed and reconstituted
when dissolved fragments re-nucleate on hydroxylated
regions.\cite{deleeuw2,criscenti,nangia1,meijer}  The dynamic equilibrium
has recently been demonstrated in Monte Carlo simulations\cite{nangia1}
using a reactive silica force field.\cite{garofalini3}

We have only observed one ring-opening reaction on each surface.  
%While
%this may be due to the finite trajectory lengths, trimer rings which
%start out containing SiOH groups have not yet been attacked by H$_2$O
%molecules, regardless of whether pK$_{\rm a}$ calculations are being
%performed on them (Fig.~\ref{fig6}).  It is possible that
%SiOH groups on trimer rings electrically and sterically hinder approach of
%H$_2$O to its Si atom.  The rupturing of such SiOH-containing cyclic
%trimers may be more efficiently investigated
%using modern reactive force fields.\cite{garofalini}  
The limited AIMD trajectory length does not conclusively allow us to predict
how many trimer rings persist at the liquid water-amorphous silica interface
as a function of time.  
Therefore we do not definitively assign this structure
to the observed pK$_{\rm a}$=4.5 SiOH group, and instead pose it as a
challenge to experimental work, including single molecule spectroscopy,\cite{ye}
to determine whether they are sufficiently abundant over time to account
for the 19\% of all silanol groups shown to exhibit high
acidity.\cite{ong,shen05}  We also point out that, while
cyclic silica trimers are well known to react with moist air, other
popular crystalline silica model surfaces should also be hydrolytically
unstable.\cite{shultz}  Thus, the hydroxylated $\alpha$-quartz (0001) and
$\beta$-cristobalite (100) surfaces, with $\sigma_{\rm SiOH}$ $\sim$
9 and 8~nm$^{-2}$ respectively, are often used as models to study the
interface between liquid water and generic silica solids in classical
MD simulations.  However,
if these simulations permit chemical reactions over long enough times,
we speculate that some of the SiOH groups on such surfaces may also react with
water\cite{shultz,nangia1} in a way to reduce the silanol surface density
towards the amorphous silica $\sigma_{\rm SiOH}$=4.6~nm$^{-2}$ observed in
experiments.\cite{iler,zhuravlev}

{\it Discussions} AIMD-based potential of mean force calculations
simulations have been demonstrated to yield reproducible pK$_{\rm a}$
for chemically equivalent silanol groups.  The statistical uncertainties
of our simulation protocol are estimated to be about 0.3-0.5~pH unit,
consistent with explicit calculations on two chemically
equivalent SiOH.   Therefore these simulations should
reliably distinguish relative pK$_{\rm a}$ of heterogeneous SiOH
groups 4 pH units apart.  Resolving hydroxyl pK$_{\rm a}$ on other
surfaces may remain a challenge if the acidities are less widely
separated.  

Our pK$_{\rm a}$ calculations suggest that comparative studies between 
SiO$_2$ and other material surfaces will be extremely interesting.  One
intriguing candidate surface is a quartz surface densely functionalized with
carboxylic acid groups.  To our knowledge, this is the only other material
surface which clearly exhibits bimodal pK$_{\rm a}$ behavior.\cite{geiger1} 
Other candidates are the different facets of crystalline alumina,
which may feature several pK$_{\rm a}$'s unresolved into distinct, measurable
components.\cite{alumina2,alumina3,alumina4}  Structural motifs such as chemical
connectivity and inter-hydroxyl hydrogen bonding have also been invoked
to explain the pK$_a$ in these systems; as mentioned in the text, inter-hydroxyl
hydrogen bonding may affect the pK$_{\rm a}$ of the more ionic Al$_2$O$_3$
surfaces more strongly than on any form of silica surfaces.  See SI, Sec.~S6,
for more details on these material systems.
Finally, SFVG spectra\cite{shen05} and time-dependent acid-base phenomena
on quartz\cite{geiger2} can also be investigated in the future.  

\section{Conclusions}
 
In this paper, we have performed AIMD pK$_{\rm a}$ calculations
on five representative crystalline silica surfaces plus a molecular
system exhibiting different silanol (SiOH) structural motifs.  From the
results, we have conclusively shown that the more acidic of the two
pK$_{\rm a}$ observed in experiments cannot, as previously
proposed,\cite{dong,lorenz,fan1,mori1,nanosilica,others,shaw2,rosenholm}
be explained by the existence of silanol groups with certain chemical
connectivities or inter-silanol hydrogen bonding. In fact, we find
pK$_{\rm a}\sim$ 4.5 silanol groups only on strained surfaces with sparse
silanol coverage.  While our demonstration of the existience of such low
pK$_{\rm a}$ SiOH groups is necessarily somewhat speculative, this study
highlights the role of defected regions as the most promising candidate
to explain the elusive bimodal acid-base behavior of silica
surfaces.\cite{ong,shen05,allen}
Assigning structural motifs to the more strongly acidic SiOH groups
is particularly crucial in non-reactive force field-based modeling of
silica nanofluidic channels, where the preferentially deprotonated SiOH
sites at neutral pH have to be assigned in a static
way.\cite{pore,lorenz,hartnig,schulten,schulten1,jctn} In the process of
studying the acid-base behavior, we also observe irreversible, water-assisted
ring-opening reactions of strained silica trimer rings in contact with
liquid water, The reaction was previously studied on wet silica
surfaces;\cite{garofalini1,lasaga,garofalini1} our AIMD simulations
demonstrate that a similar mechanism is operative at liquid water-silica
interfaces.

%\vspace*{3in}
 
%\newpage
 
\section*{Acknowledgement}
We thank Ron Shen, Steven Garofalini, Susan Rempe, Jeff Brinker,
Dave Tallant, Ying-Bing Jiang, and Franz Geiger for discussions.  This work was
supported by the Department of Energy under Contract DE-AC04-94AL85000.
Sandia is a multiprogram laboratory operated by Sandia Corporation,
a Lockheed Martin Company, for the U.S.~Deparment of Energy.
LJC acknowledge support from the U.S. DOE Office of Basic Energy Sciences,
Division of Chemical Sciences, Geosciences, and Biosciences.

\section*{Supporting Information Available}
Further information are provided regarding details of the AIMD simulations,
justification for the reaction coordinate used, quantum chemistry
calculations (including all energies and optimized geometries), 
the dynamics of hydrogen bond fluctuations on silica surfaces,
proton exchange due to multiple deprotonation, and a brief overview
of acid-base behavior in alumina and carboxylate acid functionalized
surfaces.  This information is available free of charge via the Internet
at {\tt http://pubs.acs.org/}.
 
%\clearpage
%\newpage


\begin{references}
 
\bibitem{book}
Brinker,~C.J.; Scherer,~G.W.  {\it Sol-Gel Science};
Academic Press, London, 1990, Ch. 10.
 
\bibitem{iler}
Iler,~R.K. {\it The Chemistry of Silica: Solubility, Polymerization,
Colloid and Surface Properties, and Biochemistry}; Wiley, New york, 1979.
 
\bibitem{dekker1}
Stein,~D.; Kruithof,~M.; Dekker,~C.  {\it Phys. Rev. Lett.} {\bf 2004},
{\it 93}, 035901.
 
\bibitem{baca}
Baca,~H.K.; Ashley,~C.; Carnes,~E.; Lopez,~D.; Flemming,~J.; Dunphy,~D.;
Singh,~S.; Chen,~Z.; Liu,~N.G.; Fan,~H.Y.; Lopez,~G.P.; Brozik,~S.M.; 
Werner-Washburne,~M.; Brinker,~C.J. {\it Science} {\bf 2006},
{\it 313}, 337.
 
\bibitem{yang}
Fan,~R.; Huh,~S.; Yan.~R.; Arnold,~J.; Yang,~P.D.  {\it Nature Mater.}
{\bf 2008}, {\it 7}, 303.
 
\bibitem{oil}
Kokal, S.; Tang, T.; Schramm, L.; Sayegh, S.  {\it Colloids and Surfaces
A: Physicochemical and Engineering Aspects} {\bf 1995}, {\it 94}, 253.
 
\bibitem{cpmd}
Car,~R.; Parrinello~M.  {\it Phys.~Rev.~Lett.} {\bf 1985}, {\it 55}, 2471.
 
\bibitem{sprik}
Sprik~M.  {\it Chem.~Phys.} {\bf 2000}, {\it 258}, 139.
 
\bibitem{klein}
Ivanov,~I.; Chen,~B.; Raugei,~S.; Klein,~M.L.  {\it J.~Phys.~Chem.~B}
{\bf 2006}, {\it 110}, 6365.
 
\bibitem{parrin1}
Park,~J.M.; Laio~A.; Iannuzzi,~M.; Parrinello,~M. {\it J. Am. Chem. Soc.}
{\bf 2006}, {\it 128}, 11318.
 
\bibitem{chandler}
Geissler,~P.L.; Dellago,~C.; Chandler,~D.; Hutter,~J.; Parrinello,~M.
{\it Science}, {\bf 2001}, {\it 291}, 2121.

\bibitem{pore}
Leung,~K.; Rempe,~S.B.; Lorenz,~C.D.  {\it Phys.~Rev.~Lett.}
{\bf 2006}, {\it 96}, 095504.
 
\bibitem{hass}
Hass,~K.C.; Schneider,~W.F.; Curioni,~A.; Andreoni,~W. {\it Science}
{\bf 1998}, {\it 282}, 265.
 
\bibitem{eisenthal1}
Eisenthal,~K.B. {\it Chem. Rev.} {\bf 2006}, {\it 106}, 1462.
 
\bibitem{shen1}
Shen~Y.R.; Ostroverkhov,~V. {\it Chem. Rev.} {\bf 2006}, {\bf 106}, 1140.
 
\bibitem{ong}
Ong,~S.W.; Zhao,~X.L.; Eisenthal,~K.B. 
{\it Chem. Phys. Lett.} {\bf 1992}, {\it 191}, 327.
 
\bibitem{shen05}
Ostroverkhov,~V.; Waychunas,~G.A.; Shen,~Y.R. 
{\it Phys. Rev. Lett.} {\bf 2005}, {\it 94}, 046102.
 
\bibitem{allen}
Allen,~L.H.; Matijevic,~E.; Meites,~L.
{\it J. Inorg. Nucl. Chem.} {\bf 1971}, {\bf 33}, 1293.

\bibitem{duval}
Duval,~Y.; Mielczarski,~J.A.; Pokrovsky,~O.S.; Mielczarski, E.;
Ehrhardt,~J.J. {\it J.~Phys.~Chem.~B} {\bf 2002}, {\it 106}, 2937.
 
\bibitem{shultz}
Li,~I.; Bandara,~J.; Shultz,~M.J. {\it Langmuir}, {\bf 2004}, {\it 20}, 10474.
 
\bibitem{schindler}
von Schindler, P.; Kamber,~H.R. {\it Helv. Chim. Acta}
{\bf 1968}, {\it 51}, 1781.
 
\bibitem{dong}
Dong,~Y.; Pappu,~S.V; Xu,~Z. {\it Anal. Chem.} {\bf 1998}, {\it 70}, 4730. %iso
 
\bibitem{lorenz}
Lorenz,~C.D.; Crozier,~P.S.; Anderson,~J.A.; Travesset,~A.
{\it J.~Phys.~Chem.~C} {\bf 2008}, {\it 112}, 10222.
 
\bibitem{fan1}
Fan,~H.-F.; Li,~F.P.; Zare,~R.N.; Lin,~K.-C., {\it Anal. Chem.} {\bf 2007},
{\it 79}, 3654. %iso
 
\bibitem{mori1}
Mori,~T.; Kuroda,~Y.; Yoshikawa,~Y.; Nagao,~M.; Kittaka,~S. {\it Langmuir}
{\bf 18}, 1595 (2005). % Q2
 
\bibitem{nanosilica}
Vance,~F.W.; Lemon,~B.I.; Ekhoff,~J.A.; Hupp,~J.T.
{\it J.~Phys.~Chem.~B} {\bf 1998}, {\it 102}, 1845.

\bibitem{others}
O'Reilly,~J.P.; Butts,~C.P.; I'Anson,~I.A.; Shaw,~A.M.
{\it J.~Am.~Chem.~Soc.} {\bf 2005}, {\it 127}, 1632.
 
\bibitem{shaw2}
Fisk,~J.D.; Batten,~R.; Jones,~G.; O'Reilly,~J.P.; Shaw,~A.M.
{\it J. Phys. Chem. B} {\bf 2005}, {\it 109}, 14475. % Q3
 
\bibitem{rosenholm}
Rosenholm,~J.M.; Czuryskiewicz,~T.; Kleitz,~F.; Rosenholm,~J.B.;
Linden,~M. {\it Langmuir} {\bf 2007}, {\it 23}, 4315. % Q3
 
\bibitem{zhuravlev}
Zhuravlev, L.T.  {\it Colloids Surf. A} {\bf 2000}, {\it 173}, 1.
 
\bibitem{hiemstra}
See, e.g., Hiemstra,~T.; De Wit,~J.C.M.; Van Riemsdijk,~W.H.
{\it J. Coll. Interface Sci.} {\bf 1989}, {\it 133}, 105,
and references therein.
 
\bibitem{bickmore}
Bickmore,~B.R.; Tadanier,~C.J.; Rosso,~K.M.; Monn,~W.D.; Eggett,~D.L.
{\it Geochim. Cosmochim. Acta} {\bf 2004}, {\it 68}, 2025.

\bibitem{sahai2}
Tossell,~J.A.; Sahai,~N.
{\it Geochim. Cosmochim.~Acta} {\bf 2000}, {\it 64}, 4097.

\bibitem{rustad1}
Rustad, J.R.; Dixon, D.A.; Kubicki, J.D.; Felmy, A.R.  {\it J. Phys. Chem.
A} {\bf 2000}, {\it 104}, 4051.

\bibitem{sefcik}
Sefcik, J.; Goddard, W.A. {\it Geochmim.~et Cosmochim.~Acta} {\bf 2001},
{\it 65}, 4435.
 
\bibitem{mauri}
See, e.g., Tielens,~F.; Gervais,~C.; Lambert,~J.F.; Mauri,~F.; Costa~D.
{\it Chem.~Mater.} {\bf 2008}, {\it 20}, 3336, and references therein.
 
\bibitem{marx}
Nair,~N.N.; Schreiner,~E.; Marx~D.
{\it J.~Am.~Chem.~Soc.} {\bf 2008}, {\it 130}, 14148.
%{\it J.~Am.~Chem.~Soc.} {\bf 2006}, {\it 128}, 13815.
 
\bibitem{galli2}
Cicero,~G.; Grossman,~J.C.; Schwegler,~E.; Gygi,~F.; Galli,~G.
{\it J.~Am.~Chem.~Soc.} {\bf 2008}, {\it 130}, 1871.
 
\bibitem{mundy}
Kuo,~I.-F.W.; Mundy,~C.J.  {\it Science} {\bf 2004}, {\bf 303}, 658.
 
\bibitem{car}
Kudin,~K.N.; Car,~R.  
{\it J.~Am.~Chem.~Soc.} {\bf 2008}, {\it 130}, 3915.

\bibitem{angelo}
Liu,~L.-M.; Krack,~M.; Michaelides,~A. {\it J. Chem. Phys.} {\bf 2009},
{\it 130}, 234702.
 
\bibitem{funel}
Takahara,~S.; Sumiyama,~N.; Kittaka,~S.; Yamaguchi,~T.; Bellissent-Funel~M.C.
{\it J.~Phys.~Chem.~B} {\bf 2005}, {\it 109}, 11231.
 
\bibitem{rossky}
Lee~S.H.; Rossky,~P.J. 
{\it J.~Chem.~Phys.} {\bf 1994}, {\it 100}, 3334.

\bibitem{pbe}
Perdew,~J.P.; Burke,~K.; Ernzerhof,~K.M. 
{\it Phys.~Rev.~Lett.} {\bf 1996}, {\it 77}, 3865.
 
\bibitem{vasp}
Kresse~G.; Joubert~D.  {\it Phys.~Rev.~B} {\bf 1999}, {\it 59}, 1758.
 
\bibitem{vasp1}
Kresse~G.; Furthm\"{u}ller~J. 
{\it Phys.~Rev.~B} {\bf 54}, 11169-11186 (1996).

\bibitem{charmm}
Lopes,~P.~E.~M.; Murashov,~V.; Tazi,~M.; Demchuk,~E.; MacKerell,~A.D.
{\it J.~Phys.~Chem.~B} {\bf 2006}, {\it 110}, 2782.
                                                                                
\bibitem{spce}
Berendsen,~H.J.C.; Grigera,~J.R.; Straatsma,~T.P. {\it J.~Phys.~Chem.}
{\bf 1987}, {\it 91}, 6269.
                                                                                
\bibitem{gcmc}
Martin~M.G.; Thompson,~A.P. {\it Fluid Phase Equil.} {\bf 2004}, {\it 217}, 105.

\bibitem{book1}
Chandler~D.  {\it Introduction to Modern Statistical Mechanics};
Oxford, New York, 1997, Ch.~6
 
\bibitem{klein1}
Blumberger~J.; Klein~M.~L. {\it Chem.~Phys.~Lett.} {\bf 2006}, {\it 422}, 210.
 
\bibitem{co2}
Leung,~K.; Nielsen, I.M.B.; Criscenti,~L.J. {\it J. Phys. Chem.~B} {\bf 2007},
{\it 111}, 4453.

\bibitem{meta1}
Laio, A.; Parrinello, M.  {\it Proc. Natl. Acad. Sci. USA} {\bf 2002},
{\it 99}, 12562.

\bibitem{meta2}
Iannuzzi, M.; Laio, A.; Parrinello, M. {\it Phys. Rev. Lett.} {\bf 2003},
{\it 90}, 238302.

\bibitem{meta3}
Laio, A.; Gervasio, F.L.  {\it Rep. Prog. Phys.} {\bf 2008}, {\it 71}, 126601.
 
\bibitem{g03}
Frisch,~M.~J. {\it et al.}, Gaussian 03 (Revision C.02), Gaussian Inc.,
Wallingford, CT, 2004.
 
\bibitem{lyp}
Lee~C.T.; Yang~W.T.; Parr~R.G.
{\it Phys.~Rev.~B} {\bf 1988}, {\it 37}, 785.
                                                                                
\bibitem{b3lyp}
Becke~A.~D.  {\it J.~Chem.~Phys.}, {\bf 1993}, {\it 98}, 5648.
 
\bibitem{CCSD(T)}
Raghavachari,~K.; Trucks,~G.W.; Pople,~J.A.; Head-Gordon,~M.
{\it Chem. Phys. Lett.} {\bf 1989}, {\it 157}, 479.

\bibitem{meng}
Yang, J.; Meng, S.; Xu, L.F.; Wang, E.G. {\it Phys. Rev. Lett.} {\bf 2004},
{\it 92}, 146102.
 
\bibitem{fitts}
Fitts,~J.P.; Shang,~X.M.; Flynn,~G.W.; Heinz,~T.F.; Eisenthal,~K.B.
{\it J.~Phys.~Chem.~B} {\bf 2005}; {\it 109}, 7981.

\bibitem{chuang}
Chuang, I.-S.; Maciel, G.E. {\it J. Phys. Chem. B}, {\bf 1997}, {\it 101}, 3052.

\bibitem{yb}
Preliminary experimental measurements suggest that pH of zero charge increases
inside narrow silica nanoporour membranes (Jiang, Y.-B., private
communications).

\bibitem{garofalini2}
Xu, S.; Scherer, G.W.; Mahadevan, T.S.; Garofalini, S.H.  {\it Langmuir}
{\bf 2009}, {\it 25}, 5076, and references therein.
 
\bibitem{tiss}
Tissandier,~M.D.; Cowen,~K.A.; Feng,~W.Y.; Grunlach,~E.; Cohen,~M.H.;
Earhart,~A.D.; Coe,~J.V.; Tuttle,~T.R. {\it J.~Phys. Chem.~A} {\bf 1998},
{\it 102}, 7787.
 
\bibitem{hartnig}
Hartnig, C.; Witschel, W.; Spohr, E.; Gallo, P.; Ricci, M.A.; Rovere, M.
{\it J. Mol. Liq.} {\bf 2000}, {\it 85}, 127.

\bibitem{schulten}
Cruz-Chu, E.R.; Aksimentiev, A.; Schulten, K.  {\it J. Phys. Chem. B}
{\bf 2006}, {\it 110}, 21497.

\bibitem{dipole}
Leung, K.  {\it J. Am. Chem. Soc.} {\bf 2008}, {\it 130}, 1808.

\bibitem{nangia}
Nangia, S.; Washton, M.M.; Mueller, K.T.; Kubicki, J.D.; Garrison, B.J.
{\it J. Phys. Chem. C} {\bf 2007}, {\it 111}, 5169.
 
\bibitem{deleeuw1}
Du,~Z.~M.; de Leeuw,~N.H. 
{\it Surf. Sci.} {\bf 2004}, {\it 554}, 193.
 
\bibitem{deleeuw2}
Du,~Z.~M.; de Leeuw,~N.H. 
{\it Dalton Trans.} {\bf 2006}, {\it 22}, 2623.
 
\bibitem{alumina2}
Braunschweig,~B.; Eissner,~S.; Daum,~W. 
{\it J.~Phys.~Chem.~C} {\bf 2008}, {\it 112}, 1751.
 
\bibitem{rickman}
Rickman, J.M.; Srolovitz, D.J.  {\it Surf. Sci.} {\bf 1993}, {\it 284}, 211.
 
\bibitem{strain}
Brinker,~C.J.; Kirkpatrick,~R.J.; Tallant,~D.R.; Bunker,~B.C.;
Montez,~B.  {\it J.~Non-Cryst. Solids}, {\bf 1988}, {\it 99}, 418.
 
\bibitem{wallace1}
Wallace,~S.; West,~J.K.; Hench,~L.L. {\it J. Non-Cryst. Solids}
{\it 152}, 101 (1993).

\bibitem{garofalini1}
Garofalini, S.H.  {\it J. Non-Cryst. Solids} {\bf 1990}, {\it 120}, 1.

\bibitem{garofalini}
Mahadevan~T.S.; Garofalini,~S.H. 
{\it J. Phys. Chem. C} {\bf 2008}, {\it 112}, 1507.
 
\bibitem{lasaga}
Lasaga, A.C.  {\it Rev. Mineralogy} {\bf 1990}, {\it 23}, 17.
 
\bibitem{singer}
Hassanali~A.A.; Singer,~S.J. 
{\it J.~Phys.~Chem.~B} {\bf 2007}, {\it 111}, 11181; see Fig.~9.
 
\bibitem{criscenti}
The mechanism may be the reverse of the hydrolysis pathway
examined by Criscenti, L.J.; Kubicki, J.D.; Brantley, S.L.
{\it J. Phys. Chem. A} {\bf 2006}, {\it 110}, 198.

\bibitem{nangia1}
Nangia, S.; Garrison, B.J. {\it J. Am. Chem. Soc.} {\bf 2009}, {\it 131},
9538.

\bibitem{meijer}
Trinh, T.T.; Jansen, A.P.J.; van Santen, R.A.; Meijer, E.J.  {\it J. Phys.
Chem. C} {\bf 2009}, {\it 113}, 2647.

\bibitem{garofalini3}
Feuston, B.P.; Garofalini, S.H. {\it J. Chem. Phys.} {\bf 1988}, {\it 89}, 5818.
 
\bibitem{ye}
Fu,~Y.; Collinson,~M.M.; Higgins,~D.A.  
 {\it J. Am. Chem. Soc.} {\bf 2004}, {\it 126}, 13838.
 
\bibitem{geiger1}
Konek,~C.T.; Musorrafiti,~M.J.; Al-Abadleh,~H.A.; Bertin,~P.A.;
Nguyen,~S.T.; Geiger,~F.M. 
{\it J.~Am.~Chem.~Soc.} {\bf 2004}, {\it 126}, 11754.

\bibitem{alumina3}
Zhang,~L.; Tian, C.; Waychunas, G.A.; Shen, Y.R. {\it J. Am. Chem. Soc.}
{\bf 2008}, {\it 130}, 7686.
 
\bibitem{alumina4}
Stack,~A.G.; Higgins,~S.R.; Eggleston,~C.M. 
{\it Geochim. Cosmochim.  Acta} {\bf 2001}, {\it 65}, 3055.
 
\bibitem{geiger2}
Gibbs-Davis, J.M.; Kruk, J.J.; Konek, C.T.; Scheidt, K.A.; Geiger, F.M.
{\it J.~Am.~Chem.~Soc.} {\bf 2008}, {\it 130}, 15444.

\bibitem{schulten1}
Cruz-Chu, E.R.; Aksimentiev, A.; Schulten, K. {\it J. Phys. Chem. C}
{\bf 2008}, {\it 113}, 1850.

\bibitem{jctn}
Leung, K.; Rempe, S.B. {\it J. Theor. Comput. Nanoscience} {\bf 2009}, 
{\it 6}, 1948.
{\bf 2008}, {\it 113}, 1850.
                                                                                
\end{references}
\end{document}